\documentclass{article}
\pdfoutput=1
\usepackage{amsmath}
\usepackage{amssymb}
\usepackage{graphicx}
\usepackage{multirow}
\usepackage{xcolor}

\pdfpageheight\paperheight
\pdfpagewidth\paperwidth

\usepackage{jcappub}

\renewcommand{\[}{\begin{equation}}
\renewcommand{\]}{\end{equation}} 
\usepackage{array}

\begin{document}

\title{Early Cosmology Constrained}

\author[a,b,c,d,e]{Licia Verde,}
\author[f,a]{Emilio Bellini,}
\author[g]{Cassio Pigozzo,}
\author[h]{Alan F. Heavens,}
\author[a,b,c,d]{Raul Jimenez}

\affiliation[a]{ICC, University of Barcelona, IEEC-UB, Mart\'i Franqu\`es, 1, E08028
Barcelona, Spain}
\affiliation[b]{ICREA, Pg. Lluis Companys 23, 08010 Barcelona, Spain}
\affiliation[c]{Radcliffe Institute for Advanced Study, Harvard University, MA 02138, USA}
\affiliation[d]{Institute for Theory and Computation, Harvard-Smithsonian Center for Astrophysics, 60 Garden Street, Cambridge, MA 02138, USA}
\affiliation[e]{Institute of Theoretical Astrophysics, University of Oslo, 0315 Oslo, Norway}
\affiliation[f]{University of Oxford, Denys Wilkinson Building, Keble Road, Oxford, OX1 3RH,  UK}
\affiliation[g]{Instituto de F\'isica, Universidade Federal da Bahia, Salvador, BA, Brasil}
\affiliation[h]{Imperial Centre for Inference and Cosmology (ICIC), Imperial College, Blackett Laboratory, Prince Consort Road, London SW7 2AZ, U.K.}

\emailAdd{liciaverde@icc.ub.edu; emilio.bellini@physics.ox.ac.uk; cpigozzo@ufba.br; a.heavens@imperial.ac.uk; raul.jimenez@icc.ub.edu}

\abstract{We investigate our knowledge of early universe cosmology by exploring how much additional energy density can be placed in different components beyond those in  the $\Lambda$CDM model. To do this we use a method to separate early- and late-universe information enclosed in observational data, thus markedly reducing the model-dependency of the conclusions. We find that the 95\% credibility regions for extra energy components of the early universe at recombination  are:  non-accelerating  additional fluid density parameter  $\Omega_{\rm MR} < 0.006$ and  extra radiation parameterised as extra effective neutrino species $2.3 < N_{\rm eff} < 3.2$ when imposing flatness. 
Our constraints thus show that even when analyzing the data in this largely model-independent way, the possibility of hiding extra energy components beyond $\Lambda$CDM in the early universe is seriously constrained by current observations. We also find that the standard ruler, the sound horizon at radiation drag, can be well determined in a way that does not depend on late-time Universe assumptions, but depends strongly on early-time physics and in particular on additional components that behave like radiation. We find that the standard ruler length determined in this way is $r_{\rm s} = 147.4 \pm 0.7$ Mpc if the radiation and neutrino components are standard, but the uncertainty increases by an order of magnitude when non-standard dark radiation components are allowed, to $r_{\rm s} = 150 \pm 5$ Mpc.
}
\maketitle

\section{Introduction}

The cosmic microwave background (CMB) is one of the most important cosmological probes and is 
the key observable in establishing the standard ($\Lambda$CDM) cosmological model. Lower redshift observations such as clustering of large-scale structure or  probes of  the expansion history (supernovae, baryon acoustic oscillations, cosmic chronometers), have been essential to confirm this picture and to constrain the properties of dark energy and other energy components of the Universe. 

When fitting the CMB power spectrum (temperature and polarization) one usually has to make simultaneous assumptions about the early and late cosmology, with the implication that the physics of both epochs are entwined in the resulting constraints.  Models of the late stages of the cosmological evolution  rely on less solid physical grounds than the early stages: the  physics at decoupling is well understood (atomic physics, general relativity and linearly perturbed Friedmann-Robertson-Walker metric) but the late time cosmic acceleration is not, relying, as it does, on a new ingredient such as dark energy.

Most constraints on cosmological parameters are formally model-dependent, with $\Lambda$CDM or extensions being routinely assumed.
It is therefore natural to ask: ``how much do we know about early cosmology?", ``how much of what we know about the early time depends on assumptions about the late cosmology?"  And conversely: ``how much of what we know about late cosmology depends on assumptions about early cosmology?"
These questions are key in the epoch of precision cosmology: once the cosmological parameters of a model are known at the $\sim 1$\%  level,  going beyond  parameter fitting becomes of interest and it is important to directly test the model itself. One way to achieve this is to perform model-independent analyses as much as possible.

Ref. \cite{Vonlanthen:2010cd} proposed that it is possible
to analyse the CMB in a manner which is  independent of the details of late-time cosmology. Essentially, varying the late-time physics introduces simple changes to the amplitude and angular scaling of the microwave background fluctuations, in the high angular wavenumber range where the integrated Sachs-Wolfe effect is unimportant. This approach, further developed by \cite{Audren:2012wb,Audren:2013nwa}, can be readily exploited, and Refs. \cite{Audren:2012wb,Audren:2013nwa} propose  that it avoids making assumptions about the most relevant late-cosmology effects, yielding ``pure" or ``disentangled" information on early cosmology alone.

Early-time observables are not the only window into early cosmology. In fact Ref.  \cite{Heavens:2014rja} (see also \cite{Sutherland:2012ys}) provides a measurement of the Baryon Acoustic Oscillation (BAO) scale -- the sound horizon at radiation drag, a quantity fixed by early cosmology -- from late-time observations.  The measurement itself is independent of assumptions of late-time physics, beyond the applicability of the Friedmann-Robertson-Walker metric, and makes no assumptions about early-time physics either.

Here we demonstrate how much we can know about early cosmology in a way that is independent on assumptions about the late-time physics. We use state-of-the-art CMB  observations to constrain the composition of the  early Universe and the properties of the major energy components. Then we explore how constraints on the sound horizon at radiation drag depend on early time physics assumptions. We finally combine these results with constraints on the  same sound horizon obtained  in a model-independent way from the latest probe of the Universe expansion history.

This paper is organized as follows. After presenting the methodology and the data sets in \S \ref{sec:method-data} we study, analyzing  CMB data in the traditional way, an early-dark energy model in \S\ref{sec:EDE} to report the latest constraints on the model and to explicitly show  the limitations of a model-dependent approach.  Pure information about early cosmology is obtained in  \S \ref{sec:earlycosmo}. Finally we present our results and conclusions in \S \ref{sec:concl}.

\section{Data and Methods}
\label{sec:method-data}

We use the newest Planck Collaboration data release from 2015 \cite{Adam:2015rua,Aghanim:2015xee,Ade:2015xua} which we  refer to as  ``Planck 2015". We consider low $\ell$ ($2 \leq \ell \leq 29$), temperature and polarization data (referred to as lowTEB)  and high $\ell$ ($\ge 30$) temperature (TT) and polarization (TEEE) data. In the analysis for section 3, we also include the effects on parameter constraints of  Planck lensing power spectrum reconstruction (lensing). 

Unless otherwise stated,  we use the  parameter inference code Monte Python \cite{Audren:2012wb} interfaced with the Boltzmann code CLASS \cite{2011arXiv1104.2932L, Blas:2011rf} to  generate samples from the posterior via  Monte Carlo Markov Chains  (MCMC).

We also use the model-independent measurement
of  the sound horizon at the end of radiation drag pioneered by   Ref.~\cite{Heavens:2014rja} and updated to the latest data by \cite{Heavens16}.
The sound horizon at  the redshift of radiation drag $z_{\rm d}$  is a standard ruler defined by early Universe physics:

\begin{equation}r_{\rm s}(z_{\rm d}) = \int_{z_{\rm d}}^{\infty}\frac{c_{\rm s}(z)}{H(z)}dz,
\end{equation}
where $c_{\rm s}$ is the sound speed and  $H$ the Hubble parameter. When estimated from CMB observations, $r_{\rm s}(z_{\rm d})$ is a so-called derived quantity:  $c_{\rm s}(z)$ and $H(z)$ are given in terms of other cosmological parameters within a specified cosmological model. Here, for simplicity, we adopt the notation $r_{\rm d} \equiv r_{\rm s}(z_{\rm d})$.  Ref.~\cite{Heavens:2014rja,Sutherland:2012ys} showed that the sound horizon, although a property set in the early Universe, could be measured directly from late-time data. 

As discussed in \cite{Cuesta:2015mqa}, $r_{\rm d}$ can be used as an anchor for the cosmic distance ladder (where the different rungs are supernovae type Ia, BAO etc.), the anchor being set by early-Universe observations. Its determination is indirect and therefore dependent on the  adopted assumptions about early Universe physics.  On the other hand, the cosmic distance ladder can be anchored at $z=0$ using  the local determination of $H_0$. This ``direct" ladder, when including BAO and $H_0$ measurements  yields an estimate of $r_{\rm d}$  which does not depend on assumptions about early Universe physics (but relies on the observation of  standard candles and a standard ruler).

The results of Ref. \cite{Heavens:2014rja} have been updated  by \cite{Heavens16}, obtaining
$r_{d,SBH}=141.1\pm 5.5$ Mpc (using  JLA type IA SNe data \cite{Betoule:2014frx}, BAO $D_V/r_{\rm s}$ measurements \cite{Kazin:2014qga,Beutler:2011hx,Cuesta:2015mqa} and the Hubble constant $H_0$ determination of Ref.\cite{Riess:2016jrr}) and $r_{d,CSB}=150.0\pm 4.7$ Mpc when  including, instead of $H_0$, cosmic chronometer measurements from passive elliptical galaxy ages \cite{Moresco:2016mzx}, yielding $H(z)$. These measurements are used to importance sample the Planck 2015 chains and obtain combined constraints.

When performing parameter inference from CMB data in the standard way, simultaneous assumptions about early- and late-time physics must be made. As a consequence, in the resulting  constraints, early and late-time physics, which should be independent, are inextricably entwined.    In \S \ref{sec:EDE},  we demonstrate this point  using a popular model for early dark energy. This motivates us to analyze CMB data   in a way that is independent  of --or  at least robust to detailed assumptions about-- late-time cosmology. 
 
\section{A practical example: Early Dark Energy}
\label{sec:EDE}
\begin{figure}[!t]
\centering
\includegraphics[height=3.7 in,width=0.7\columnwidth]{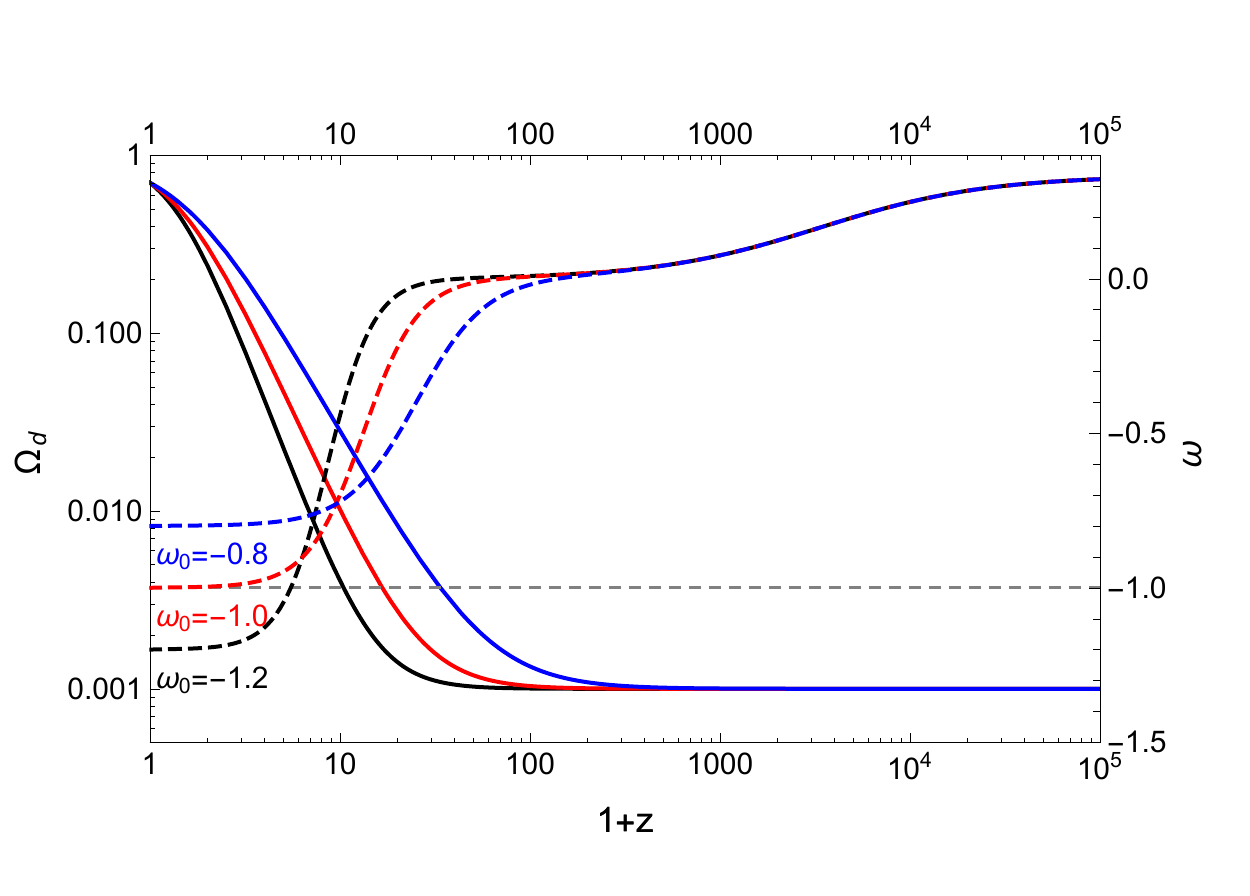}
\caption{The evolution of the  dark energy density  parameter $\Omega_{\rm d}(z)$ [\ref{EDE}] (solid lines, left axis) and of the  behaviour of the global equation of state parameter $w(z)$ from [\ref{eos}] (dashed lines, right axis). All curves were obtained for $\Omega_{\rm m,0}=0.3$, $\Omega_{\rm d,0}\approx 0.7$, $\Omega_{\rm d,e}=0.001$ and $z_{\rm eq}=3570$. Varying the values of  $\Omega_{\rm d,0}$ and $\Omega_{\rm d,e}$ would only change the upper and lower limits of $\Omega_{\rm d}(z)$ (dotted grey lines). Three examples are showed for different values of $w_0$: $-1.2$ (black curves), $-1$ (red curves) and $-0.8$ (blue curves). $w(z)$ goes from a positive value during the radiation era, evolving smoothly to $w_0$, and different values of $w_0$ make the transition from $\Omega_{\rm d,e}$ to $\Omega_{\rm d,0}$ sharper or longer.}
\label{DoranPlot}
\end{figure}

\begin{figure}[!t]
\centering
\includegraphics[height=3.6 in,width=0.7\columnwidth]{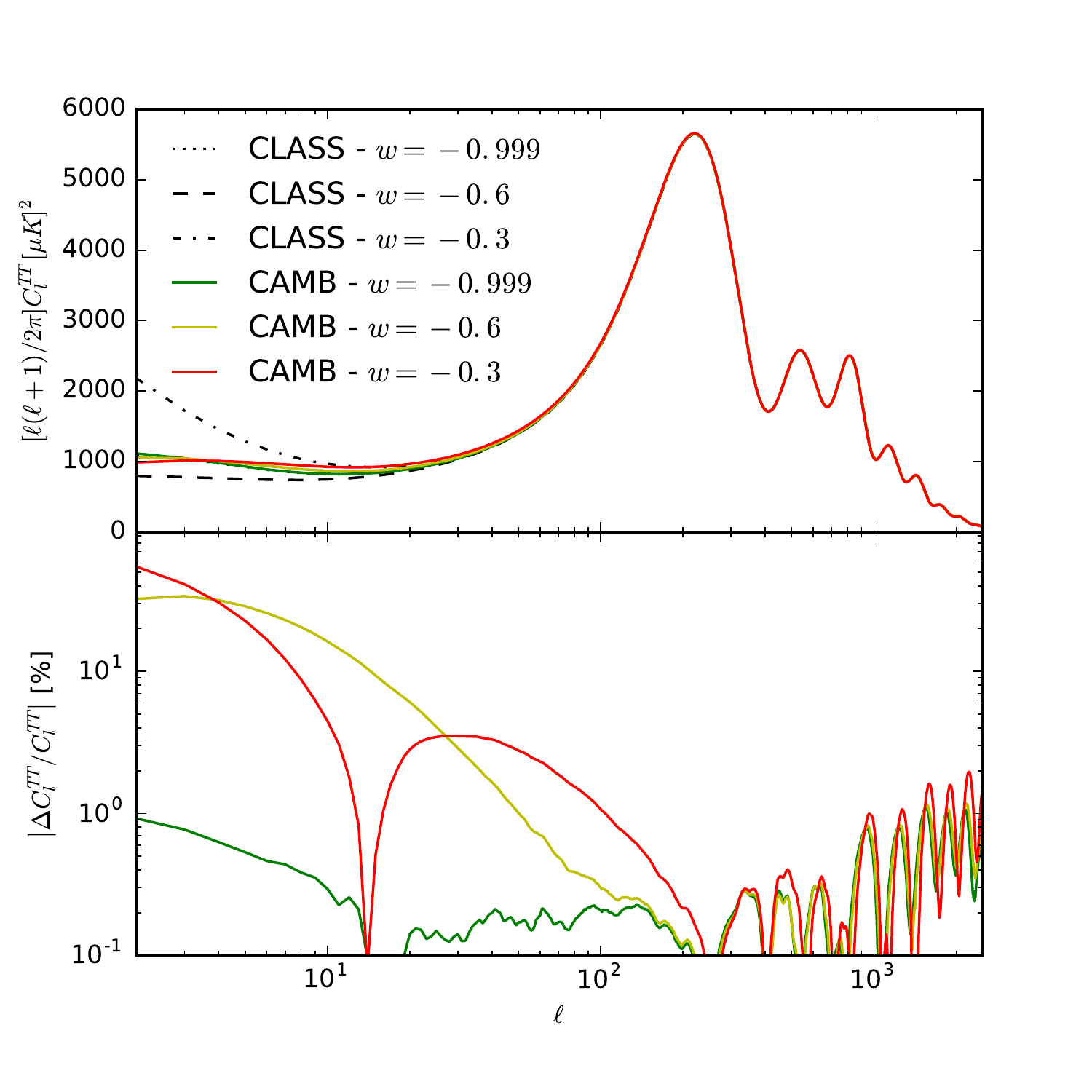}
\includegraphics[height=3.6 in,width=0.7\columnwidth]{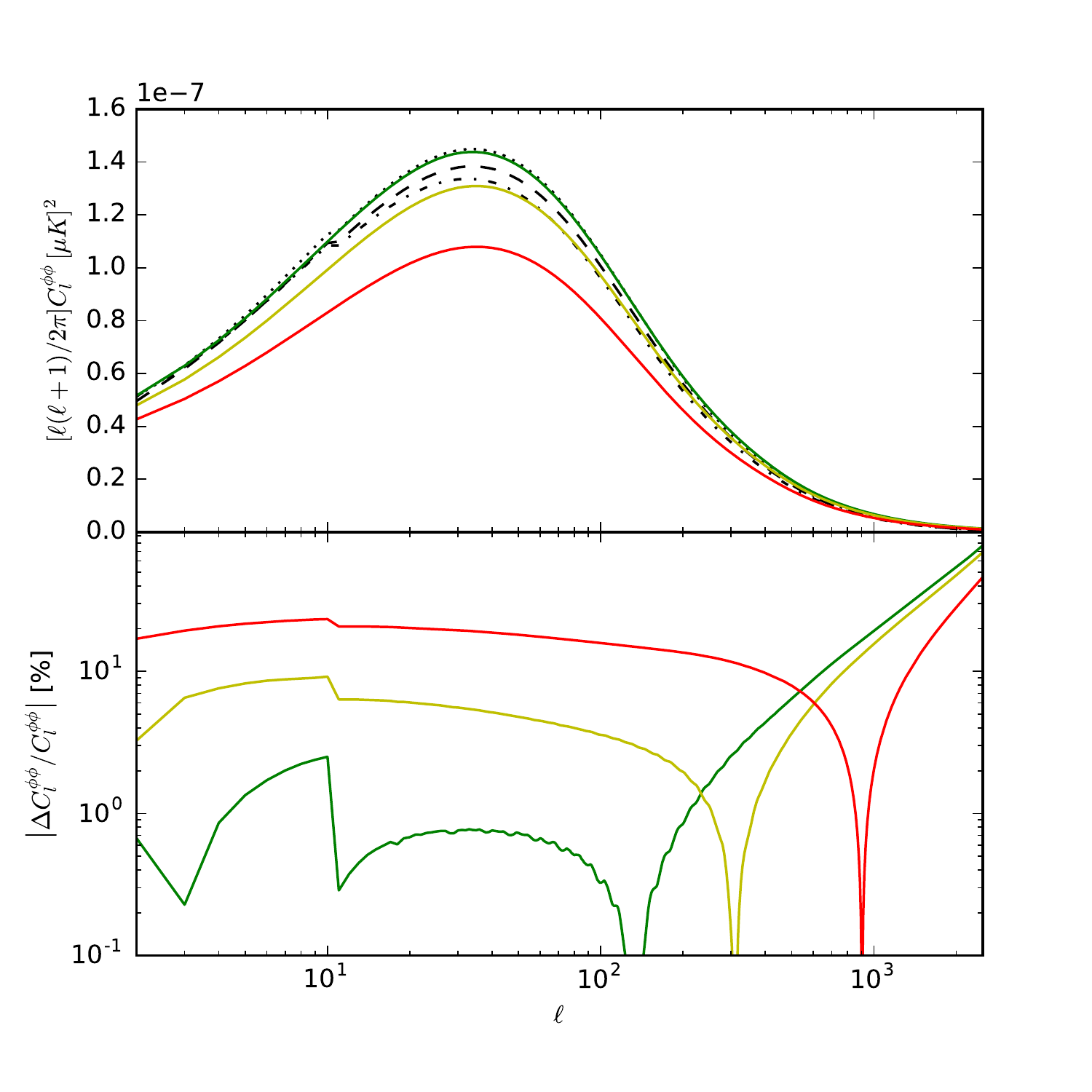}
\caption{Temperature (upper panel) and  lensing (lower panel) power spectra for different values of the equation of state parameter $w_0$, keeping fixed all the other cosmological parameters to a fiducial EDE model. For comparison, we show the results obtained by ``CLASS'' (background only) and ``CAMB'' (with perturbations). The discontinuity observed in the $C_{\ell}^{\phi\phi}$ (bottom panel) at $\ell=10$ is due to the shift between  the implementation of the Limber approximation at high $\ell$ and the full calculation at low $\ell$ in the Boltzmann code. The presence of this sharp feature does not affect the likelihood calculation in any significant way.}
\label{class_vs_camb}
\end{figure}

 We begin with a model-dependent analysis for Early Dark Energy (EDE), which we make model-independent in \S\ref{sec:earlycosmo}. While in standard $\Lambda$CDM the contribution of dark energy at early times is fully negligible, dynamical dark energy models might yield a non-negligible contribution at high redshifts. In particular such models may be similar to $\Lambda$CDM at late times (thus being compatible with late-time geometric probes such as SNe) but have significant dark energy density early on and as such they go under the name of early dark energy (EDE) models \cite{Wetterich:2004pv}. EDE slows the growth of structures at early times and changes the heights and positions of the acoustic peaks in the CMB power spectrum.
A widely used model is the one proposed by \cite{Doran:2006kp}, which assumes a constant fraction of early dark energy $\Omega_{\rm d,e}$ until a transition at recent times. This model  has the advantage of requiring only two extra parameters compared to $\Lambda$CDM. It has been used extensively and in particular it was explored in the official analysis of the Planck 2013 \cite{Ade:2013sjv} and Planck 2015 data \cite{Ade:2015rim}.

Instead of parameterising the  dark energy equation of state, $w(a)$, the authors of \cite{Doran:2006kp} proposed a parametrization of the dark energy density as a function of scale factor $\Omega_{\rm d}(a)$. Thus the Friedmann equation for a flat universe can be rewritten as

\begin{equation}
\frac{H^2(a)}{H_0^2} = \frac{\Omega_{\rm m,0}a^{-3}+\Omega_{\rm rel,0}a^{-4}}{1-\Omega_{\rm d}(a)},
\end{equation}
where $\Omega_{\rm rel,0}$ is the  density  parameter  of relativistic  species (i.e., neutrinos and photons) extrapolated to the present day, and $\Omega_{\rm m,0}$ is the present-day matter density parameter (baryonic plus dark matter). 

The parametrization is given in terms of the present-day value of the dark energy parameter $\Omega_{\rm d,0}$, the dark energy content in the early Universe $\Omega_{\rm d,e}$ and the present-time dark energy equation of state parameter $w_0$,

\begin{equation}\label{EDE}
\Omega_{\rm d}(a)=\frac{\Omega_{\rm d,0}-\Omega_{\rm d,e}(1-a^{-3w_0})}{\Omega_{\rm d,0}+\Omega_{\rm m,0}a^{3w_0}}+\Omega_{\rm d,e}(1-a^{-3w_0}).
\end{equation}
The time evolution of the equation of state can be derived from the conservation equation of the dark energy fluid \cite{Wetterich:2004pv}
\begin{equation}
\left[ 3 w-\frac{a_{\rm eq}}{a+a_{\rm eq}} \right]\Omega_{\rm d} (1-\Omega_{\rm d})=-\frac{d\Omega_{\rm d}}{d \ln{a}},
\label{eos}
\end{equation}
were $a_{\rm eq}= \Omega_{\rm rel,0}/\Omega_{\rm m,0}$ is the scale factor at the equality between radiation and matter. The evolution of both $\Omega_{\rm d}(z)$ and $w(z)$ are represented in Figure \ref{DoranPlot} for selected  sets of parameters. It should be noted that, even if this model is classified as EDE, it has a tracking behaviour. Indeed, during radiation domination the equation of state approaches $1/3$; during matter domination it approaches $0$ and only at late times it behaves as a cosmological constant if $w_0=-1$. It is possible to verify that the dark energy density described by (\ref{EDE}) varies smoothly from $\Omega_{d,e}$ to $\Omega_{d,0}$, and the redshift of transition  is determined by 
the model's parameters $\Omega_{d,e}$, $\Omega_{d,0}$ and $w_0$.

According to \cite{Doran:2006kp} this parameterization gives a  monotonic function for $\Omega_{\rm d}(z)$, as long as 
\begin{equation}
\Omega_{\rm d,e} \lesssim \frac{\Omega_{\rm d,0}}{2-\Omega_{\rm d,0}}.
\end{equation}

Contrary to the standard $\Lambda$CDM, this model has an evolving dark energy density. This means that it has to be interpreted as a parametrization for a dynamical degree of freedom (e.g.\ a scalar field) whose scope is to modify the expansion history of the Universe. However, any scalar field has fluctuations, and their contribution in principle should be considered when evolving the perturbations. This has been done in e.g.\ \cite{Ade:2015rim} using the Boltzmann code CAMB \cite{Lewis:1999bs} and considering a parametrized quintessence fluid at the perturbative level, i.e.\ a fluid with sound speed $c_{\rm s}^2=1$ and no anisotropic stress. Then it is clear that the differences between this model and a cosmological constant can come from two different sectors, i.e.\ {\it i)} a different expansion history and {\it ii)} a pure perturbative effect characterized by the additional degree of freedom. The choice of a quintessence field is arbitrary, but a relativistic sound speed ensures that DE perturbations are suppressed on sub-horizon scales and the only effect that potentially modifies the evolution of the perturbations is a different expansion history.

The presence of dark energy perturbations  couples early-time and late-time constraints: perturbations in the dark energy component are most relevant at late-time but their effect is seen in the CMB. To illustrate this point we also consider another approach where we   take into account only the contribution given by the different expansion history and not by the perturbed quintessence field using a modification of the public code ``CLASS".  The effect can be appreciated  in Fig.\ \ref{class_vs_camb}, where we show the temperature and lensing power spectra for different values of $w_0$. ``CAMB'' lines are obtained considering perturbations of a quintessence field, while for the ``CLASS'' lines we modified just the background evolution. Since the effects of a quintessence field could be relevant on very large scales (close to the cosmological horizon), it is not surprising to see that  the temperature power spectrum is affected mostly at low $\ell$, while the inclusion of perturbations modifies the gravitational potential  power spectrum $C_{\ell}^{\phi\phi}$ uniformly. 

The main motivation to consider just a modified expansion history is that the standard equations for the evolution of a minimally coupled scalar field (as quintessence) are singular when the equation of state approaches $w=-1$ (and thus a hard prior in $w> -1$ has to be imposed). This is a well known property of simple single scalar field models (e.g.\ \cite{Ballesteros:2010ks}), but it is not true for more complicated theories. Then, the ``background only" approach (hereafter, CLASS implementation) ensures that we can cross the phantom divide and explore models  with $w<-1$. 

So while  the  Planck Collaboration, when dealing with Eq. ~\ref{EDE}, could only obtain upper limits for the relevant parameters, adopting hard priors $\Omega_{\rm d,e}\geq 0$ and $w_0> -1$ \cite{Ade:2015rim}, we explore,  for the first time, the region where $w_0<-1$ and $\Omega_{\rm d,e}<0$.\footnote{Of course for models where the density parameter  goes negative this component cannot be interpreted as a fluid.}

\begin{figure}[!htb]
\centering
\includegraphics[width=\textwidth]{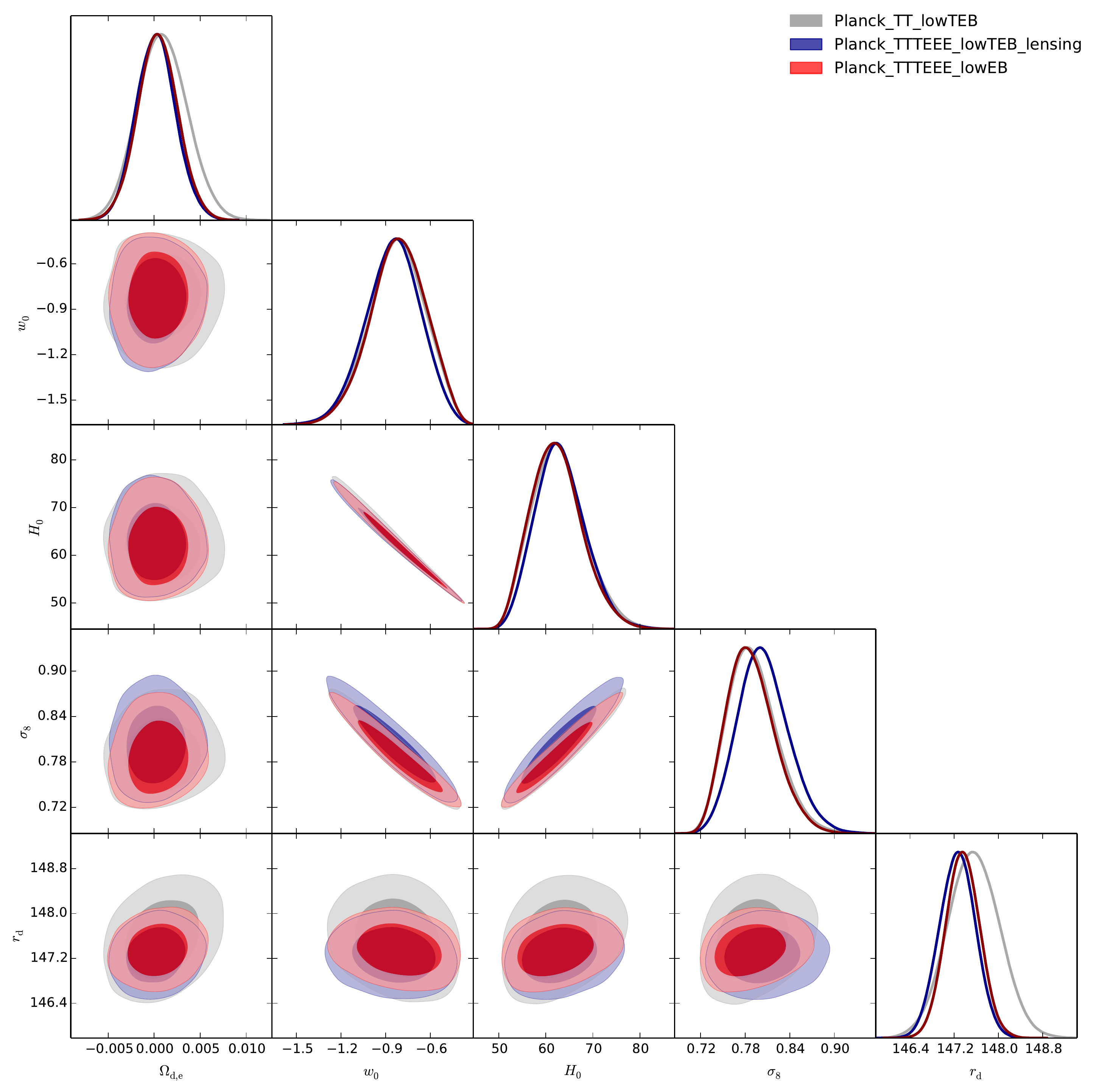}
\caption{Marginalized distributions of selected cosmological parameters for the early dark energy model of Sec. 2  accounting only for contributions to the background (``CLASS implementation"). Only  Planck 2015 data are considered for different combinations of likelihoods: high-$\ell$ temperature plus lensing; high-$\ell$ temperature and polarization; and high-$\ell$ temperature and polarization plus lensing.}
\label{fig:EDE1nopertplanckonly}
\end{figure}

\begin{figure}[!htb]
\centering
\includegraphics[width=\textwidth]{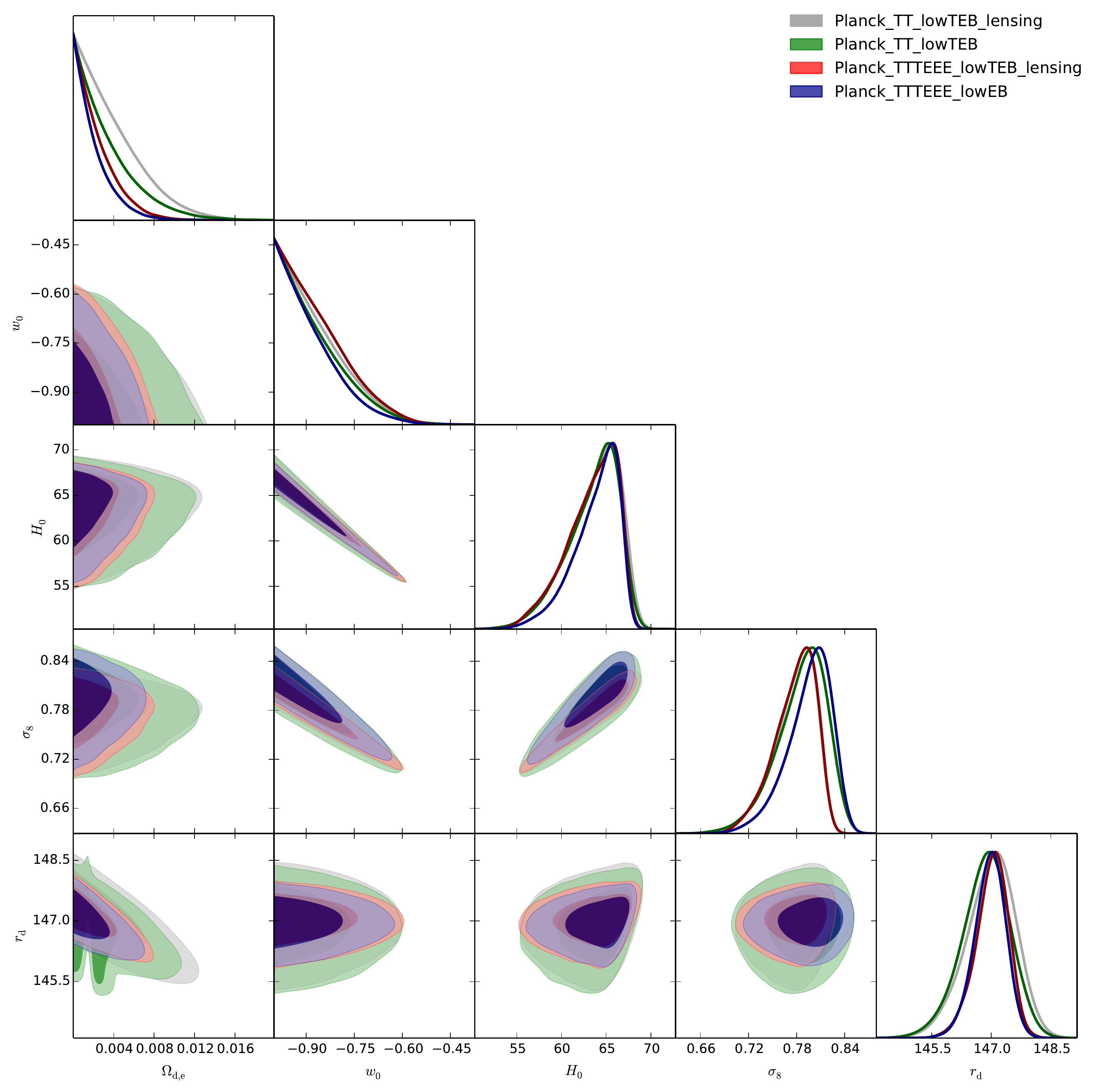}
\caption{Marginalized distributions of selected cosmological parameters for the early dark energy models of Sec. 2 accounting also for the effect of perturbations (``CAMB implementation"). Only Planck   2015 data are considered for different combinations of likelihoods: high-$\ell$ temperature plus lensing; high-$\ell$ temperature and polarization; and high-$\ell$ temperature and polarization plus lensing. The apparent sharp ``cliff" in the $H_0$-$r_{\rm s}$ plane is due to the prior $w_0>-1$.}
\label{fig:EDE1pertplanckonly}
\end{figure}

In Figs.~\ref{fig:EDE1nopertplanckonly} and  ~\ref{fig:EDE1pertplanckonly} we  compare the cosmological  constraints  obtained with the two approaches.
In Fig.~\ref{fig:EDE1nopertplanckonly} we show the marginalized posterior constraints for this early dark energy model  accounting only for contributions to the background (CLASS implementation). Different combinations of Planck likelihoods (with or without high-$\ell$ polarization and  with or without lensing) are considered.  In Fig.~\ref{fig:EDE1pertplanckonly}  the  results of the analysis  including the effects of perturbations (CAMB implementation) is shown. This can be compared directly with the analysis performed by the Planck team. 

We note that there are differences in the constraints on this model between the CLASS (background only) implementation  and the CAMB (with perturbations) implementation.  Besides the fact that the CLASS implementation crosses the phantom divide and constrains values of $w_0<-1$, $\Omega_{\rm d,e}<0$, note that, even in the common region ($w_0>-1$, $\Omega_{\rm d,e}>0$)  there are differences. In particular in Fig.\ \ref{fig:EDE1pertplanckonly} (with perturbations) there is a correlation and degeneracy between $r_{\rm d}$ and $\Omega_{\rm DE}$ which is not present in Fig.\ \ref{fig:EDE1nopertplanckonly} (background-only implementation). Larger values of early dark energy are allowed when perturbations are included.

 Given the fact that so little is known about the nature of dark energy, each of the two approaches should be  considered as a phenomenological description of dark energy, each of them  capturing different physics. For example, in  the non-perturbative case, the effect of dark energy  on the background  is captured but it is assumed that perturbations  in the new degree of freedom have no observable effects. In the perturbative case  a very particular perfect fluid, with sound speed equal to 1, is considered.

As shown in Fig.\ \ref{class_vs_camb}, besides the effect on the gravitational potential power spectrum -- which affects the CMB lensing signal --   differences between the two implementations are evident at low $\ell$, and arise through the Integrated Sachs-Wolfe (ISW) effect. The ISW effect is a late-time effect, which depends on physics at redshifts much below the last scattering surface, yet it is affected by an early-time quantity: the amount of early dark energy. The same can be said about the CMB lensing signal, which arises from late-time physical effects (see \cite{KMJ}). The constraints on early dark energy so obtained are therefore highly model-dependent: they depend on assumptions of physics in the late-time Universe.

\begin{table*}
\begin{center}
\begin{tabular}{|c | c | c | c | c | l|}
\hline    
Parameter &  TT & TT\_lens & TTTEEE & TTTEEE\_lens & Code \\ \hline 

\multirow{3}{*}{$10^3\Omega_{\rm d,e}$}  & $ < 9.240$ &$<9.80$ &$< 5.53$ & $<6.25$&  CAMB\\[1ex]
  & --   & $0.9 _{-5.1} ^{ +5.4}$ & $ 0.2_{-4.2 } ^{+4.4 }$  &$0.4 _{-4.3 } ^{+4.4 }$ &  CLASS\\[1ex] \hline

\multirow{3}{*}{$w_0$} & $< -0.69$ & $<-0.68$& $<-0.72$& $<-0.68$&  CAMB\\[1ex]
 & --  & $ -0.82_{-0.36} ^{ +0.36}$ & $ -0.85_{-0.37} ^{+ 0.34}$ & $-0.82 _{ -0.36} ^{+0.36 }$ &  CLASS\\[1ex] \hline

\multirow{3}{*}{$\sigma_8$} &   $0.786^{+0.057}_{-0.065}$& $0.778^{+0.046}_{-0.055}$& $0.795^{+0.051}_{-0.060}$& $0.778^{+0.046}_{-0.055}$&  CAMB\\[1ex]
      & --  & $ 0.790_{-0.063 } ^{ +0.067}$ &  $ 0.805_{-0.067 } ^{+0.069 }$ &  $ 0.788_{-0.060} ^{+ 0.065}$ &  CLASS\\[1ex] \hline

\multirow{3}{*}{$r_{\rm d}$ [Mpc]} &  $ 146.9^{+1.2}_{-1.0}$& $147.0^{+1.2}_{-1.3}$& $146.96^{+0.73}_{-0.84}$& $147.01^{+0.80}_{-0.88}$&  CAMB\\[1ex]
      & --  & $ 147.55_{-0.92 } ^{ +0.93}$ & $147.26 _{-0.64 } ^{+0.63 }$ & $147.36 _{-0.59 } ^{+0.60 }$ &  CLASS\\[1ex] \hline

\multirow{3}{*}{$H_0$ [km s$^{-1}$ Mpc$^{-1}$]} &  $63.5 _{-6.1 }^{+4.9 }$
&  $63.6 _{-6.2 }^{+5.0 }$ & $63.7 _{-5.6 }^{+4.3 }$ & $63.4 _{-6.0 }^{+4.8 }$ &  CAMB\\[1ex]
      & --  & $62.8 _{-10.8 }^{+11.1 }$ & $63.1 _{-10.3 }^{+10.6 }$ & $62.3 _{-10.4 }^{+10.7 }$ &  CLASS\\[1ex] \hline
\end{tabular}
\label{tabruler}
\caption{Mean and 95\% credible regions for cosmological parameters with only Planck likelihoods (in different combinations).} 
\end{center}
\end{table*}

\section{Separating  Early Cosmology from Late Cosmology}
\label{sec:earlycosmo}

In the previous section we provided constraints by fitting an EDE model against recent CMB observations. Clearly, the results we presented are model dependent. But, assuming that the model we choose is sufficiently generic, one could na\"\i vely expect that our results can be legitimately interpreted as ``early times constraints". This is not correct, since our EDE model fixes the entire evolution of the universe and the physics of the CMB as we observe it depends on both early and late time cosmology. In other words, the model we chose gives predictions also for the late time expansion history, and the CMB contains this information. It is then clear that, in order to decouple early from late time physics, two steps are needed. The first one is to assume a model to describe cosmology at early times that has to be as general as possible (this point will be described in greater detail in the next section). The second step is to decouple early from late time physics in the CMB spectra.
\begin{figure}[!t]
\centering
\includegraphics[height=3.6 in,width=0.8\columnwidth]{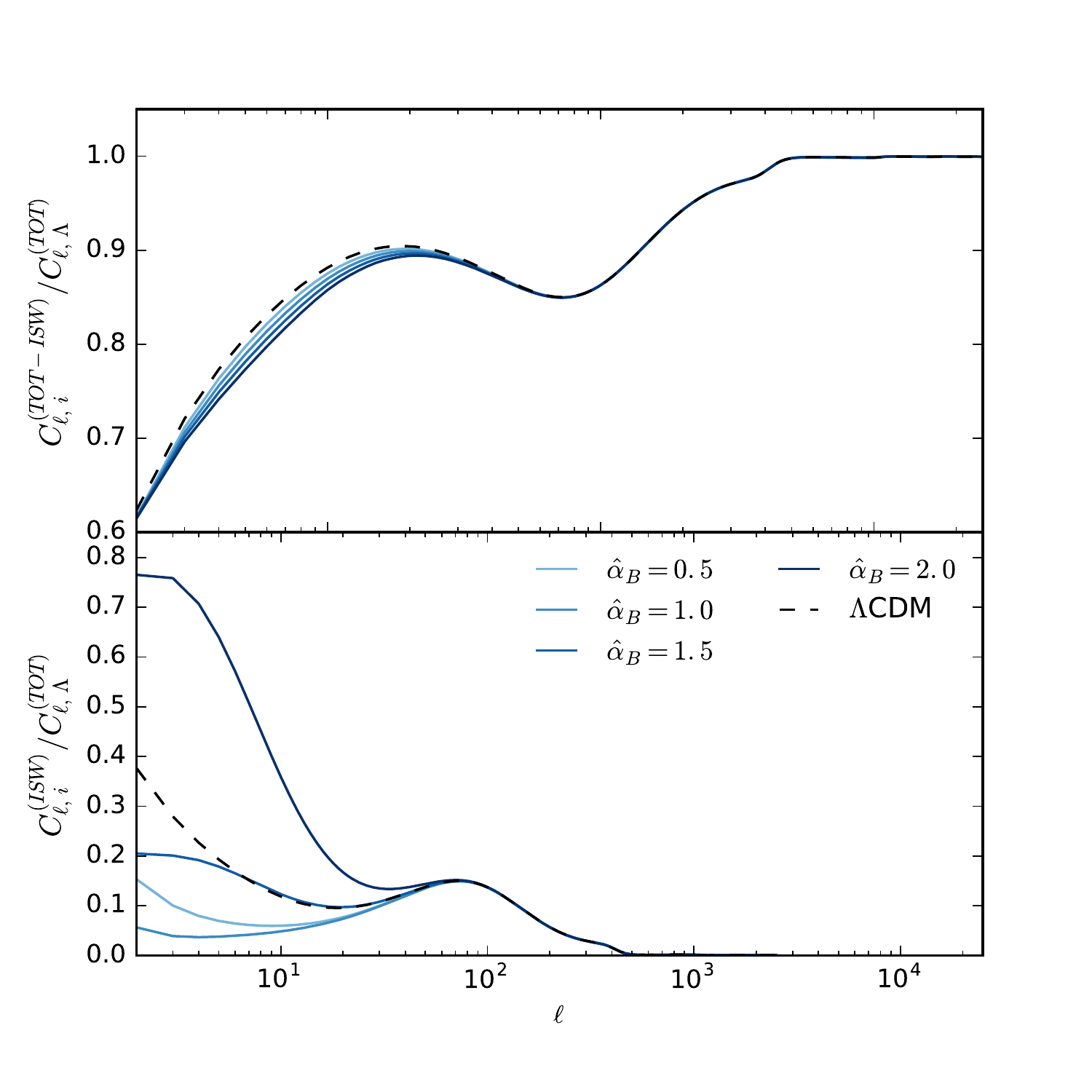}
\includegraphics[height=3.6 in,width=0.8\columnwidth]{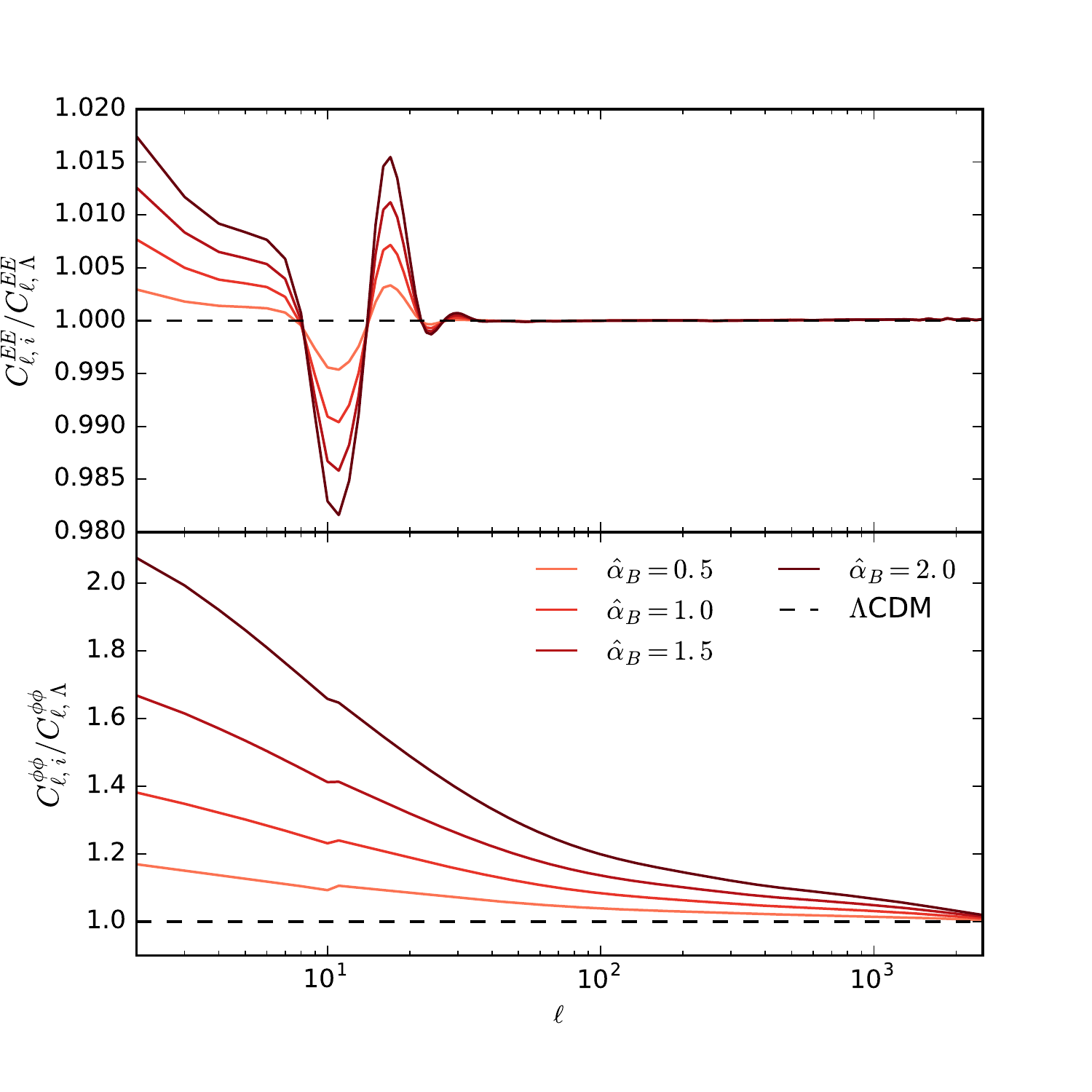}
\caption{CMB spectra (TT, EE and $\phi\phi$) for different models with the same early time evolution but different late time evolution relative to a reference model (a fiducial  standard  $\Lambda$CDM with Planck's best fit parameters).  Plots are generated with the public code \texttt{hi\_class} \cite{Zumalacarregui:2016pph}.
Models are obtained from the  fiducial $\Lambda$CDM varying  just one parameter, i.e.\ the braiding (see \cite{Bellini:2014fua} for details).
 The upper top panel shows the temperature power spectrum without the ISW effect compared to the total temperature power spectrum of our fiducial model, while the lower top panel shows just the ISW contribution to the total temperature PS. The bottom upper/lower panels show the EE/$\phi\phi$  relative power spectrum for the same models.}
\label{cl_TT_EETE}
\end{figure}
\begin{figure}[!t]
\centering
\includegraphics[height=3.75 in,width=0.8\columnwidth]{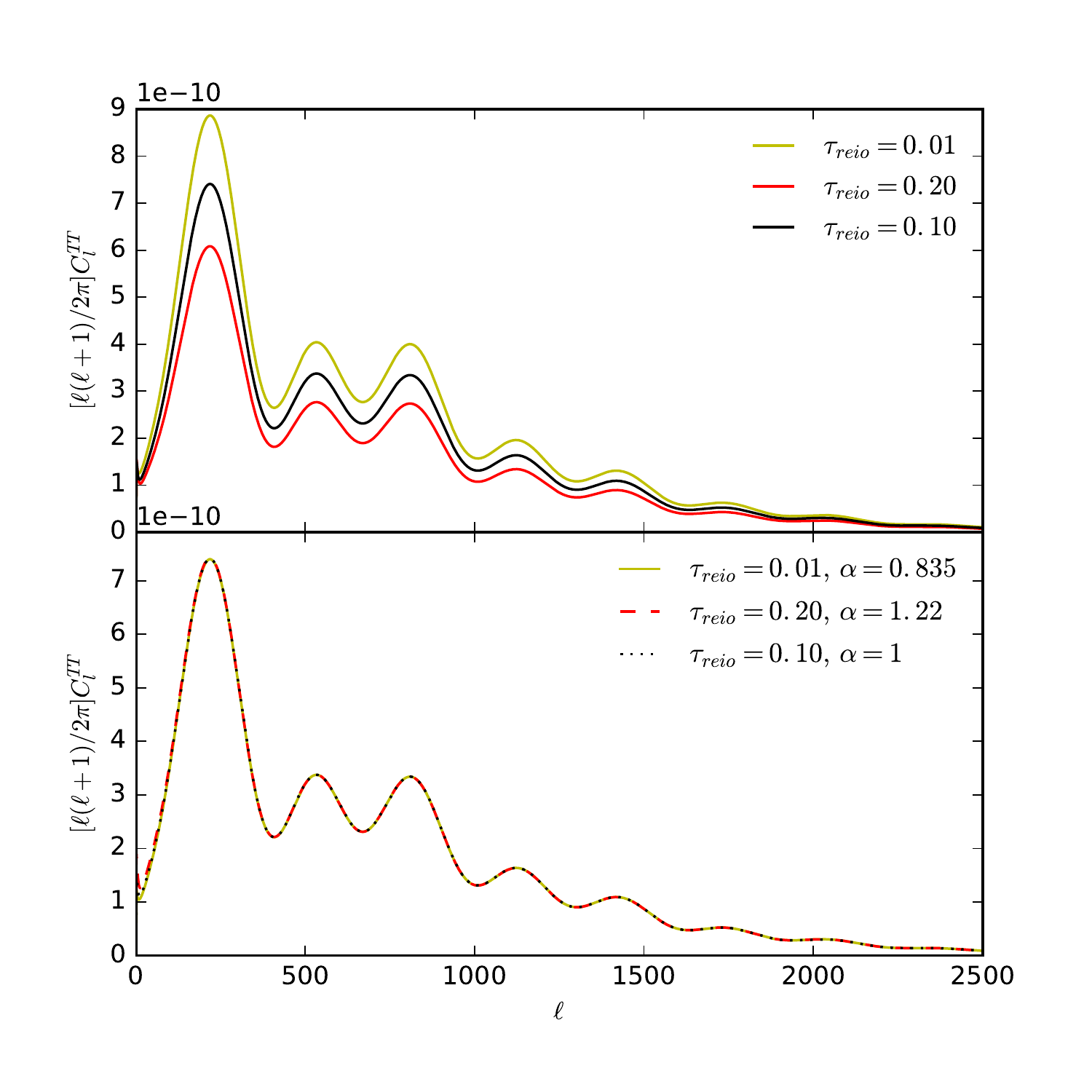}
\includegraphics[height=3.75 in,width=0.8\columnwidth]{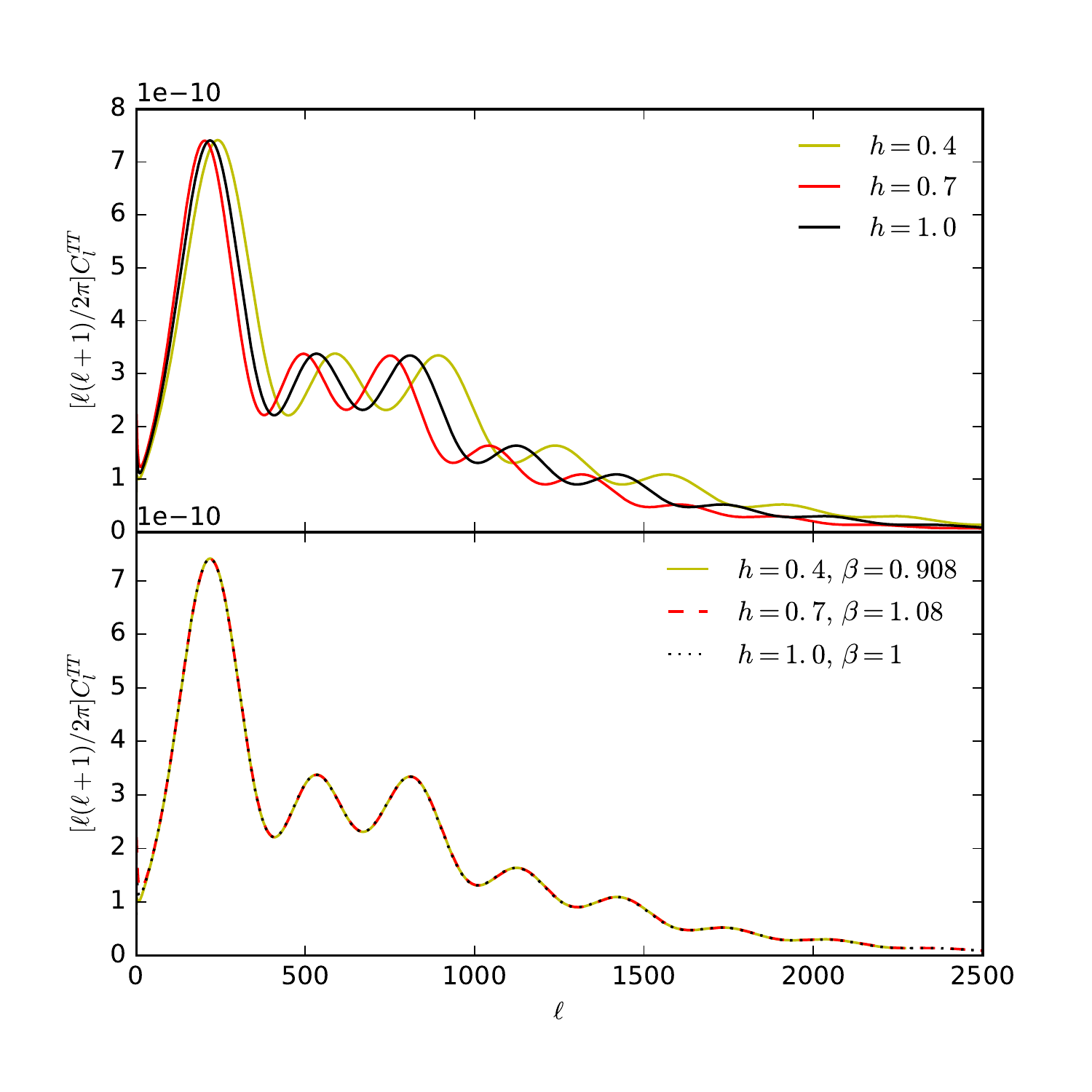}
\caption{Effects on the TT angular power spectrum due to variations of the optical depth $\tau_{reio}$ (top panels) and the Hubble parameter $h$ (bottom panels).  It is possible to note that at $\ell > 30$, variations of $\tau_{reio}$ produce a rescaling  $C_\ell^{TT}\rightarrow \alpha C_\ell^{TT}$, while variations of $h$ produce a shift of the $\ell$'s, i.e.\ $C_\ell^{TT}\rightarrow C_{\beta\ell}^{TT}$. Lower panels  show the  rescaled $C_{\ell}$ after adjusting the rescaling/shift with  appropriate values of $\alpha$ and $\beta$.}
\label{cl_rescale_shift}
\end{figure}

The key to progress is that late-time physics affects the CMB power spectrum in simple ways at high $\ell$.
\cite{Vonlanthen:2010cd,Audren:2012wb} show that the effects are limited to: (i) projection effects from real space to harmonic space, controlled by the angular diameter distance at the recombination epoch, i.e.\ $d_A\left(\tau_{rec}\right)$; (ii) the late ISW effect, affecting only small $\ell$; (iii) reionization, suppressing equally all multipoles at $\ell \gtrsim 30$. It is possible to show \cite{Vonlanthen:2010cd,Audren:2012wb} that all these effects produce just a rescaled amplitude ($C_\ell\rightarrow \alpha C_\ell$) and position ($C_\ell\rightarrow C_{\beta\ell}$) in the CMB high-$\ell$ multipoles. Then, to remove the dependence  on the late-time cosmology  we must  not only ignore low $\ell$ multipoles and  take into account (and marginalize over) the degeneracy between a direct rescaling of the amplitude and position of the CMB multipoles and late time cosmology. In addition, one has to remove the effects of late times appearing in the lensing signal. It has been argued \cite{Vonlanthen:2010cd,Audren:2012wb} that this can be done by marginalizing over an overall amplitude and tilt of the Newtonian potential. The resulting approach that we follow is:
\begin{itemize}
 \item we neglect low multipoles for temperature and polarization spectra. We will then consider only ``high-$\ell$" Planck data, i.e.\ $\ell \geq 30$; this effect is illustrated in the top  panel of Fig.\ \ref{cl_TT_EETE}, where it is shown how removing these low-$\ell$ multiples suppresses drastically the influence of late-time effects;
 \item we fix the optical depth $\tau$ to a typical value, i.e.\ $\tau = 0.01$, since any value of $\tau$ can be compensated by changing $A_{\rm s}$;
 \item we do not interpret the parameters $A_{\rm s}$ and $h$ as the scalar amplitude of fluctuations and the present-day Hubble parameter. Here $h$ effectively sets the last-scattering distance and is only connected with the current expansion rate if we assume a model connecting early to late times.  For $A_{\rm s}$ we interpret only the combination $e^{-2\tau}\, A_{\rm s}$, and for $h$ interpret only $d_A\left(\tau_{rec}\right)$, which represent vertical and horizontal scale factors of the CMB spectra. In Fig.\ \ref{cl_rescale_shift} we show that it is possible to compensate the effects of a variation of $\tau$ and $h$ with suitable rescaling and shift of the temperature $C_\ell$;
 \item we marginalize over lensing by rescaling the Newtonian potential as
 $$
 \phi\left(k,z\right) \longrightarrow A_{lp} {\left(\frac{k}{k_{lp}}\right)}^{n_{lp}} \phi\left(k,z\right)\,,
  $$
  where ($A_{lp}$, $n_{lp}$) are two new free parameters and we fixed $k_{lp}=0.1\,h$ Mpc$^{-1}$.
\end{itemize}

 In Fig.\ \ref{cl_TT_EETE} we show how the CMB power spectra are modified by choosing different dark energy/modified gravity models with the same early-time evolution but with different late-time physics. We employ the \texttt{hi\_class} \cite{Zumalacarregui:2016pph} code, which implements Horndeski's theory for dark energy/modified gravity models; Horndeski is the most general scalar-tensor theory described by second-order equations of motion, and contains many well-known dark energy/modified gravity models as special cases (see \cite{BelliniSawicki} for more details).  It can be seen (upper panels) that the temperature power spectrum is affected mostly by a change in the ISW effect. In the bottom  panels we show the polarization power spectrum (EE) and the behaviour of the lensing potential ($\phi\phi$). As expected, the polarization power spectrum is affected mostly at low multipoles, while the lensing changes at all $\ell$ (even if the effect is larger at low-$\ell$).
 As it will be clear below, in the simple implementation described in \cite{Vonlanthen:2010cd,Audren:2012wb} and adopted here,  this approach removes the late-time information for models whose late-time evolution is not too far away from a vanilla $\Lambda$CDM model. It is not a problem for models with small or vanishing dark energy. However, for models with very large values of the dark energy density parameter (roughly equivalent to $\Omega_\Lambda > 0.8$ at $z\sim 0$), the adopted priors matter and the procedure  is not guaranteed to remove all the late-time signal. In fact, as we discuss below, in these cases we find residual late-time effects in the form of ISW effects up to $\ell \sim {\cal O}(100)$. For such models, little can be said about dark energy in a model-independent way. Constraints on other quantities that are sub-dominant at late time are indeed model-independent.

Finally, as explained in \cite{Sutherland:2012ys,Heavens:2014rja} low redshift observations can be analyzed so that they yield a cosmology-independent estimate of $r_{\rm d}$, an early-time quantity.  Below we will use this constraint in combination with the above analyses to help constrain early-time physics.

\subsection{Testing Early Cosmology}
Here we concentrate on developing a set of  models generic enough to test the components  of the early Universe and their properties.
 The idea is to give as much freedom as possible to the early-time evolution of the Universe with simple and well-motivated extensions of the standard cosmological model. Then, the models under consideration are
\begin{itemize}
 \item ``LCDM": here we use the standard 6-parameter cosmological model with a cosmological constant. Even if this model cannot be considered very general, it is instructive because here $\Lambda$ cannot be interpreted as the usual cosmological constant that causes the accelerated expansion of the Universe at late times. In principle it should be   considered as an early cosmological constant decoupled from the late one;
 \item Dark energy fluid ,``DE fld": in this case we replace $\Lambda$ with a standard perfect fluid. We fix the sound speed of the scalar field to 1 and we impose a hard bound on the equation of state parameter, i.e. $w<0$ in order to avoid degeneracy with matter. The freedom here is represented by the fact that the additional scalar field is not constrained to have $w=-1$. While $w$ is a free parameter (hereafter $w^0_{\rm DE}$) it is considered constant in time;
 \item ``k": here we add to the cosmological constant $\Lambda$ an arbitrary spatial curvature with density parameter $\Omega_{\rm k}$. The idea of having a spatial curvature is often used in the literature, and it is very much constrained to be close to 0 by late-time observations (e.g.\ \cite{Ade:2015xua}. However, here we focus just on the early-time evolution and thus expect poor constraints on this parameter. It is useful to introduce it since the additional freedom given by a parameter that scales differently from a cosmological constant is able to modify substantially the early-time expansion history of the Universe;
 \item ``nu": This is effectively a standard $\Lambda$CDM model but with the addition of  non-standard neutrinos. The idea \cite{Hu:1998tk} is to modify the properties of neutrinos by varying two parameters, an effective sound speed $c_{\rm eff}^2$ and a viscosity parameter $c_{\rm vis}^2$. This case is reduced to the standard one by fixing both values to 1/3. In this model we fix one massive neutrino with mass $m=0.06eV$ but we let vary the number of massless neutrinos, i.e. $N_{\rm ur}\simeq 2$ for the standard case. The massive and massless neutrinos share the same phenomenological properties, i.e.\ $c_{\rm eff}^2$ and $c_{\rm vis}^2$ are in common. It is possible to show that non-standard neutrinos behave as a standard scalar field that scales as radiation at the background level by setting $(c_{\rm eff}^2,c_{\rm vis}^2)=(1,0)$. Then, in this model the contribution of an additional scalar field at the perturbation level can come directly from the neutrino sector at the price of renouncing the standard neutrino behaviour;
 \item ``matteriation": This is similar to the ``DE fld'' model, but with an equation of state parameter  that can take values intermediate between that of pressure-less non-relativistic matter $w=0$ and that of radiation $w=1/3$. For simplicity here we assume that this fluid contributes to a modification of the expansion history only, while the perturbations are the standard matter/metric perturbations. This case complements the ``nu" case above.
\end{itemize}

\begin{figure}
\centering
\includegraphics[width=.7\textwidth]{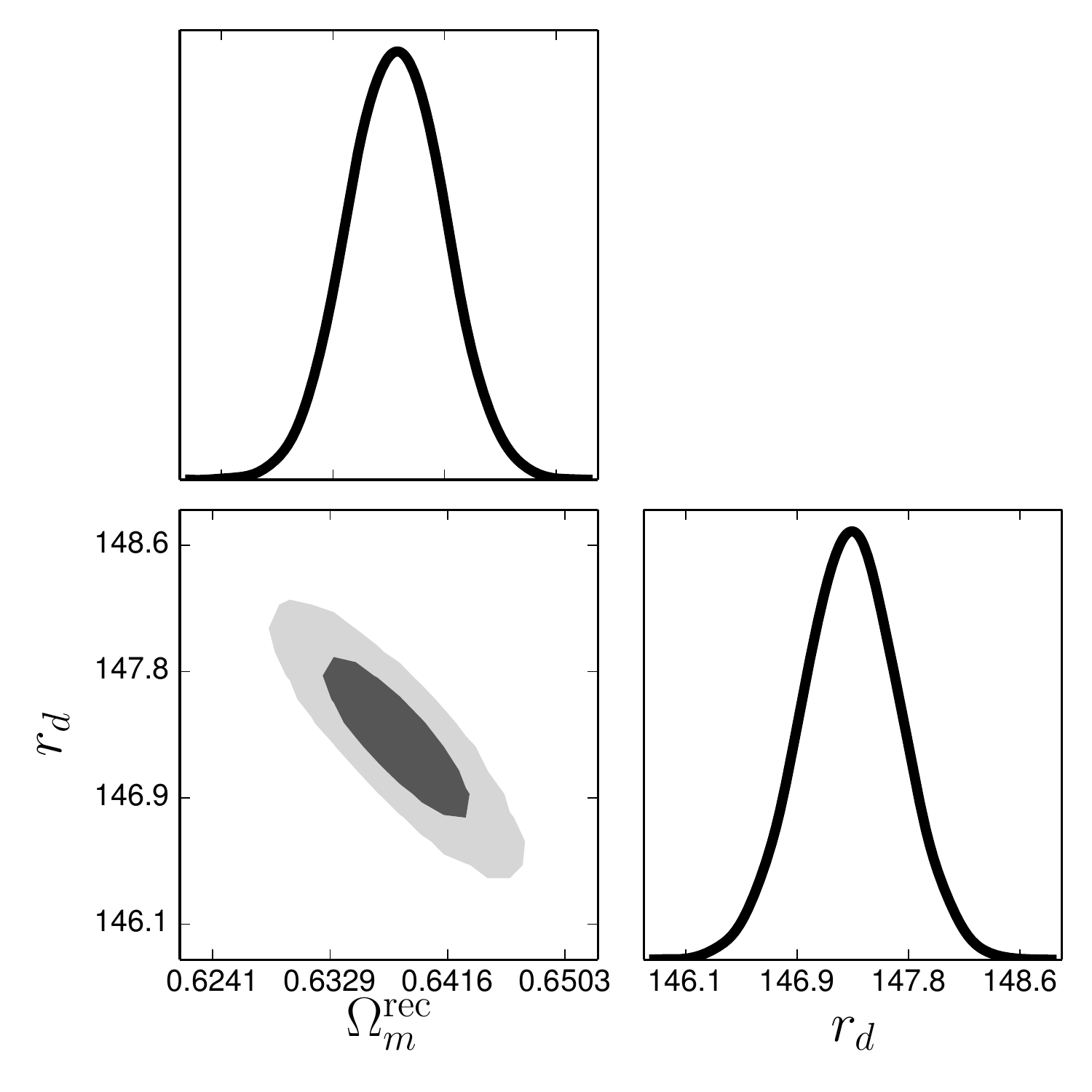}
\caption{Posterior distribution for the ``LCDM" model.\label{lambda_posterior}}
\end{figure}

\begin{figure}
\centering
\includegraphics[width=.8\textwidth]{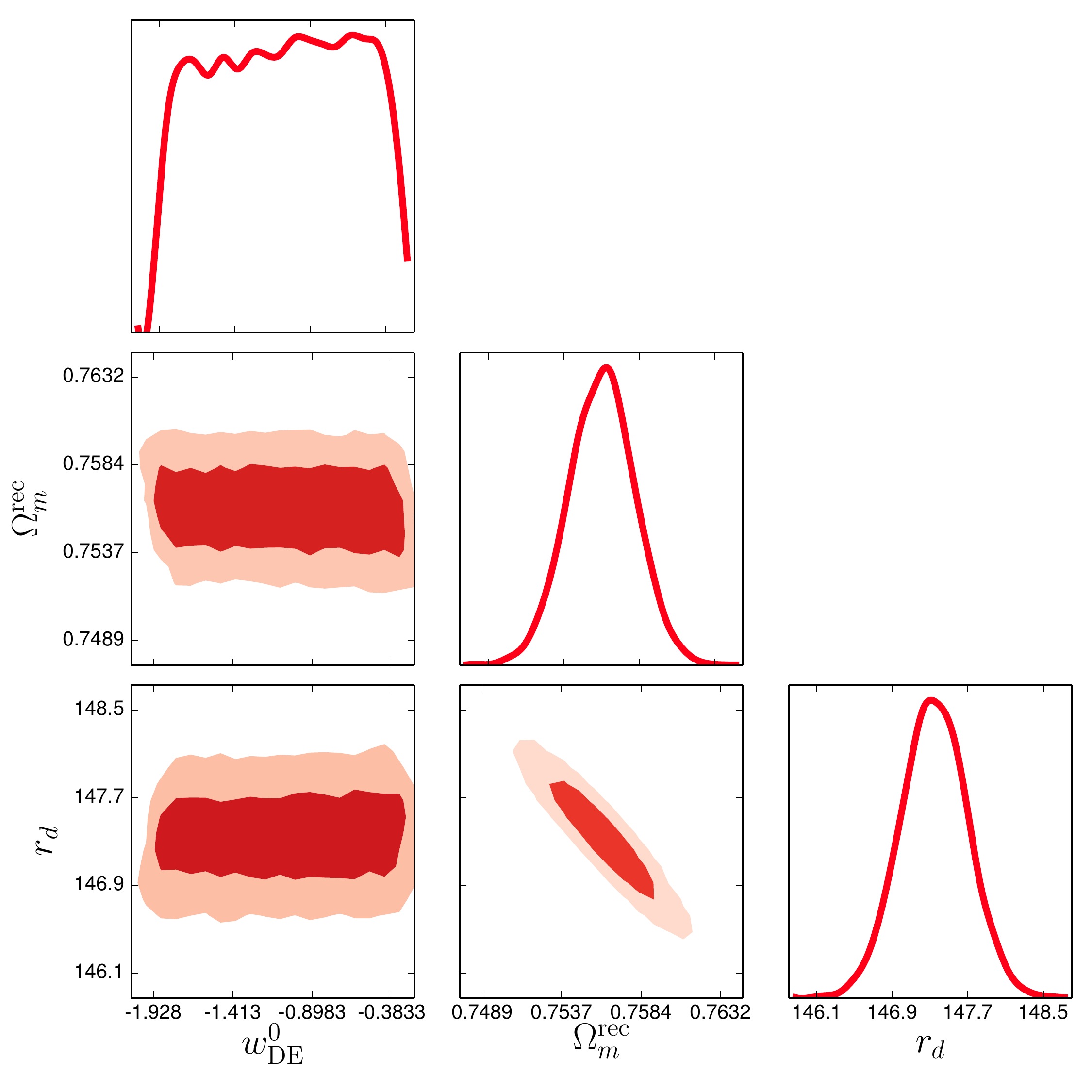}
\caption{Posterior distribution for the ``DE fld" model. \label{fld_posterior}}
\end{figure}

\begin{figure}
\centering
\includegraphics[width=.8\textwidth]{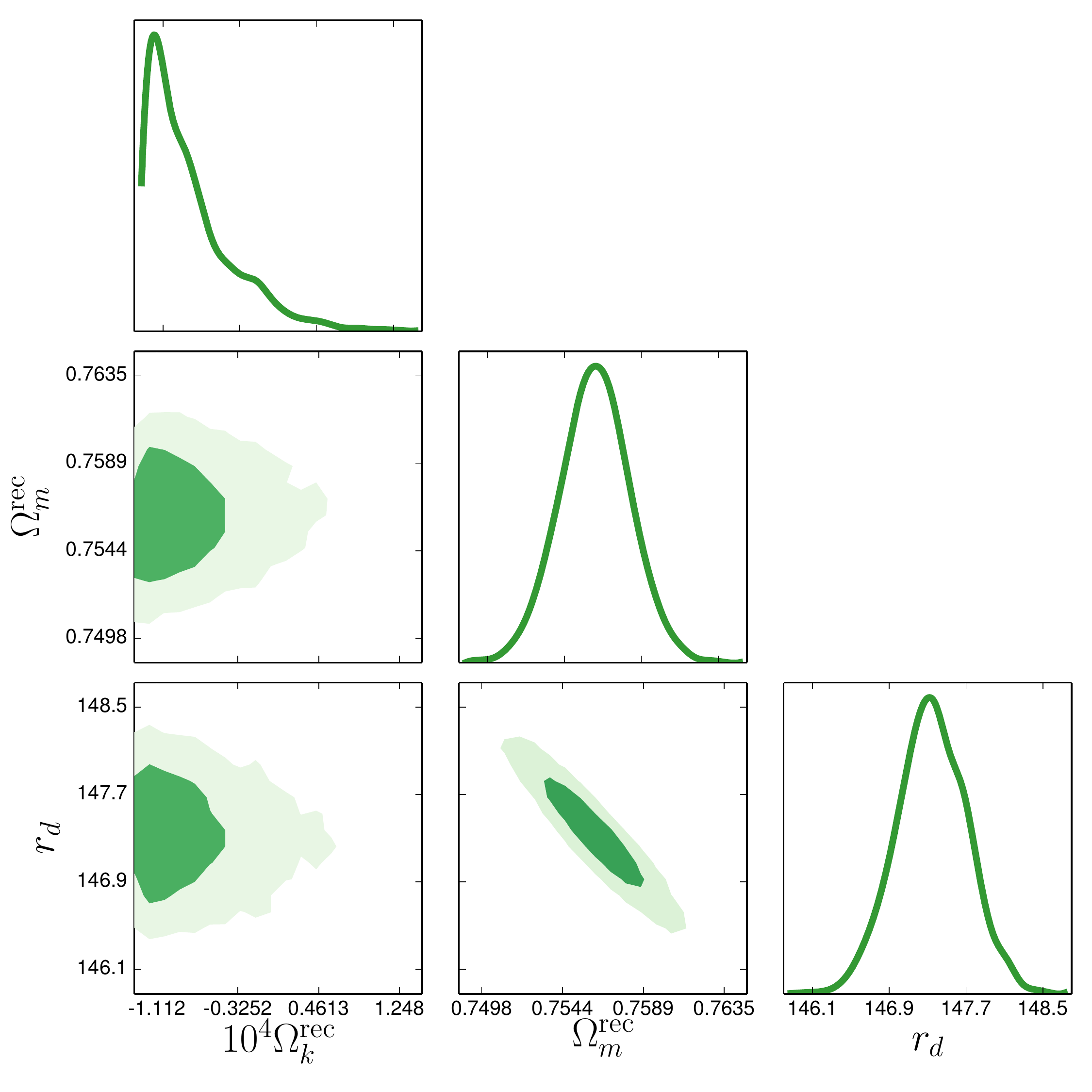}
\caption{Posterior distribution for the ``k" model.}
\label{k_posterior}
\end{figure}

\begin{figure}
\centering
\includegraphics[width=\textwidth]{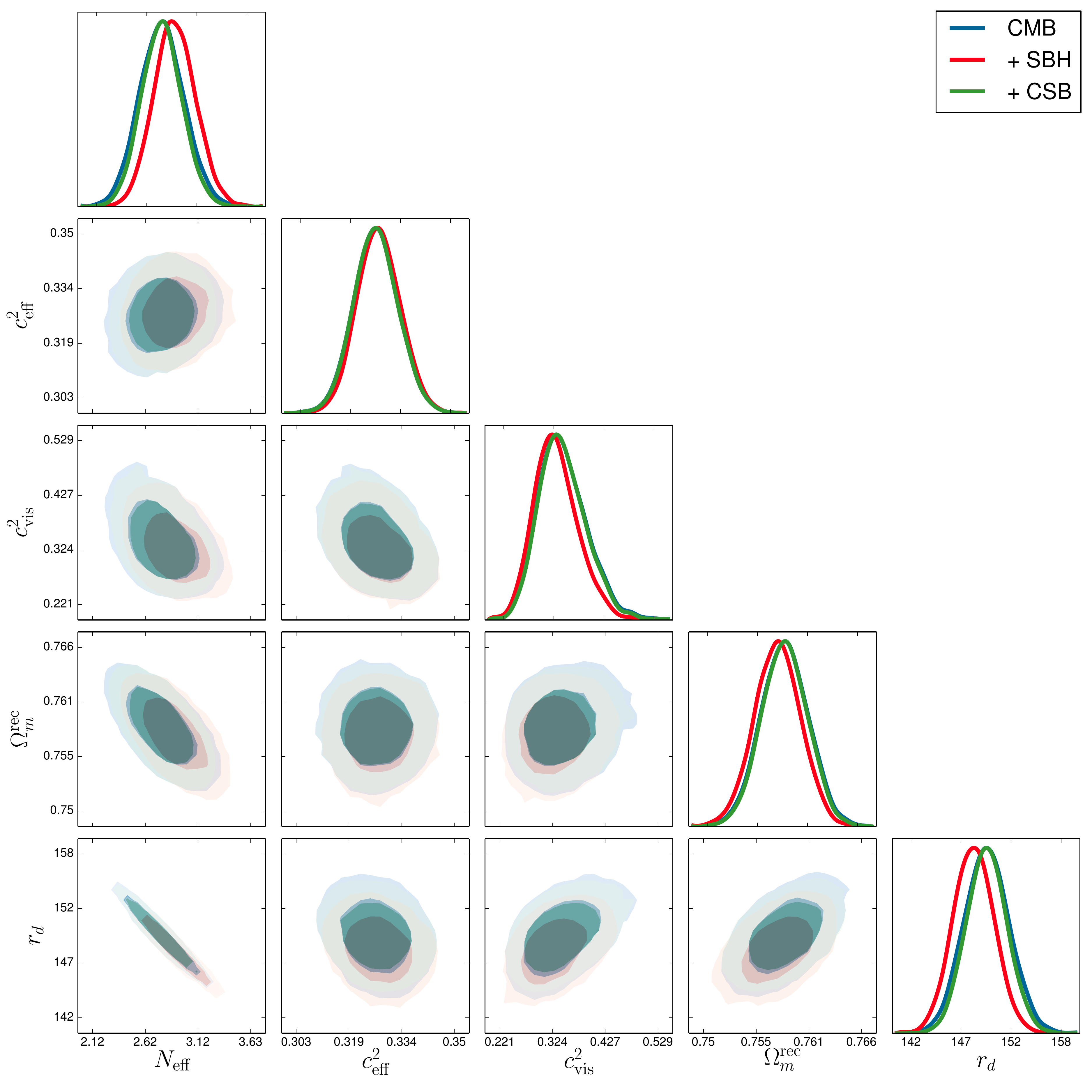}
\caption{Posterior distribution for the ``nu" model. Blue contours show the constraints without measurements of $r_{\rm s}(z_{\rm d})$. Red contours represent the constraints using the SBH dataset , i.e.\ $r_{\rm s}(z_{\rm d})=141.0\pm 5.5$. Green contours represent the constraints using the CSB dataset , i.e.\ $r_{\rm s}(z_{\rm d})=150.0\pm 4.7$.\label{nu_posterior}}
\end{figure}
\begin{figure}[!t]
\centering
\includegraphics[width=\textwidth]{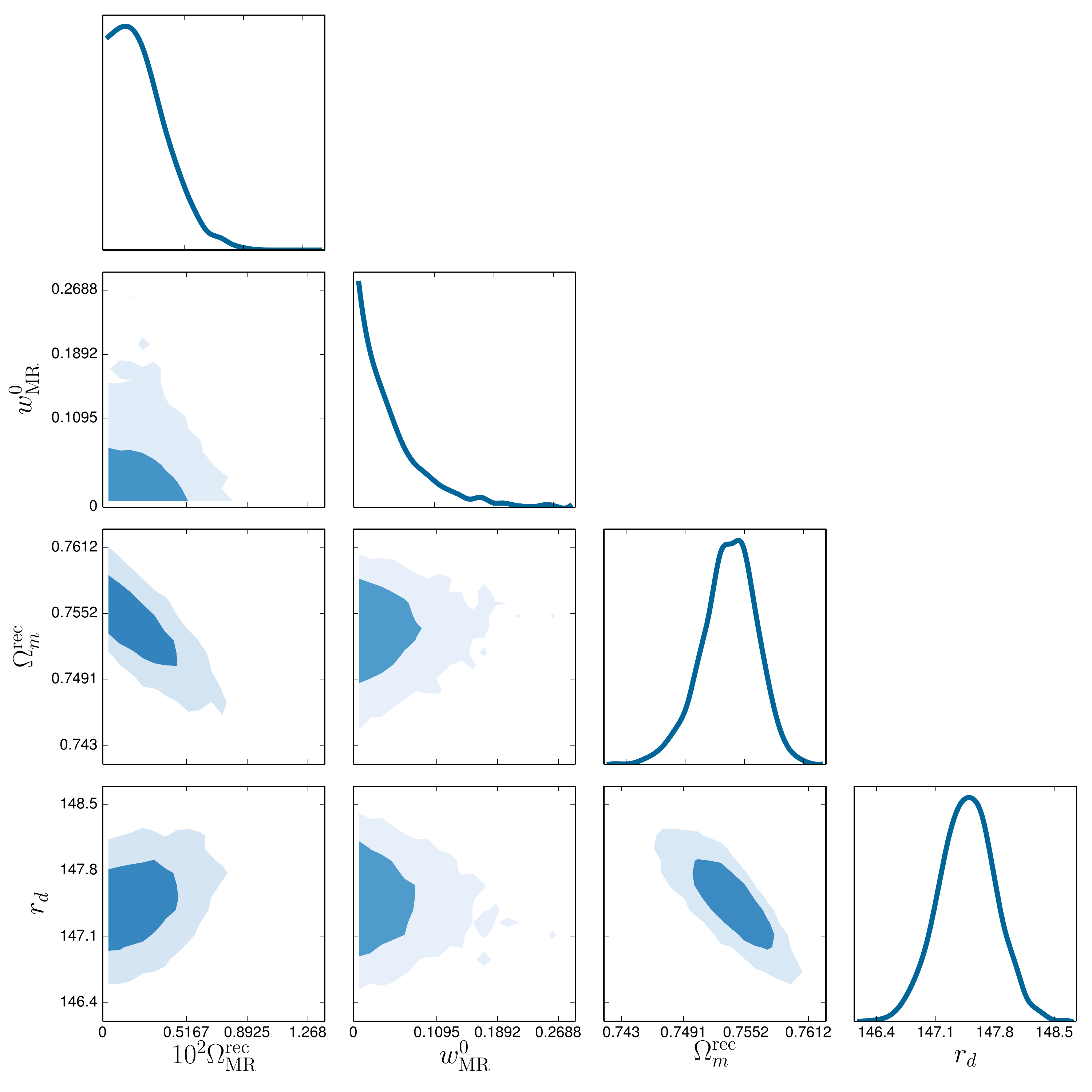}
\caption{Posterior distribution for the ``matteriation" model.\label{matteriation_posterior}}
\end{figure}

Figures \ref{lambda_posterior}, \ref{fld_posterior}, \ref{k_posterior}, \ref{nu_posterior} and \ref{matteriation_posterior} show the 68\% and 95\% joint  posterior credibility regions for the relevant parameters for the ``LCDM'', ``DE fld'', ``k'',``nu'' and ``matteriation''  cases respectively  using Planck 2015 data (TT and TEEE). Since we wish to obtain constraints on early-time cosmology only, the density parameters in the plots are reported at recombination ($z_{rec}\simeq 1090$). In all cases the densities of non-standard species are constrained to be very small compared to the matter density ($\Omega_{\rm m}\simeq 1$). 
 Special care must be made when interpreting the constraints on $\Omega_\Lambda$  especially for the curvature case.  $\Omega_\Lambda$ and  $\Omega_{\rm k}$ are degenerate but the degeneracy is not perfect. Positive values of $\Omega_{\rm k}$ are bounded from above due to the prior that  $\Omega_{\rm m}$ (as well as all other densities) must be positive and a minimal amount of $\Omega_{\rm m}$ is needed to produce the acoustic peaks. However the degeneracy   does not appear to extend to positive values beyond $\Omega_{\rm k}^{\rm rec} = 1.4\times 10^{-4}$ and  $\Omega^{\rm rec}_\Lambda$ does not extend beyond $3\times 10^{-9}$, which essentially restricts the present-day dark energy density parameter to be no more than $\sim 0.8$.  Larger values may introduce ISW fluctuations on scales where our analysis assumes there are none, so we caution against over-interpreting this limit.  Constraints on this parameter are expected to be of the order of  or below $\sim  1\%$ of the matter density at recombination \cite{Karwal:2016vyq}. 
The unphysical constraint on $\Omega_{\Lambda}$ is due to residual late-time constraints.
We note that in the non-flat case  the constraint on   $\Omega_{\Lambda}$ is much weaker  ($\sim 100\%$) than in the other cases. This does not invalidate the derived constraints on the other parameters, as $\Omega_{\Lambda}$ and $\Omega_{\rm DE}$ are in any case sub-dominant at recombination. Our findings confirm  that the Universe at early times is well described by the standard cosmological model, and there is no much room for additional components that may be relevant at early times.

 What the observations constrain remarkably well about the  early cosmology is the expansion rate $H$ at recombination (especially for standard radiation and neutrinos contributions). This  can be appreciated in Tab.~\ref{tab:models} (third line, $H_{\rm rec}$ entry). As described in \cite{ZZ} the faster the Universe is expanding at recombination, the more difficult it is for the hydrogen to recombine, increasing the ionization fraction. Larger ionization fraction yields a broader visibility function. This in turn has two effects on observable quantities. On the  CMB temperature it leads to a larger (Silk) damping of small scales anisotropies: the first acoustic peak is unaffected but  higher peaks are damped.  On the polarization, because of the increased photon mean path,  large-scale polarization anisotropies ($\ell \lesssim 800$) are enhanced.   
Planck was forecasted to constrain  $H_{\rm rec}$ at the 0.9\% level \cite{ZZ} and  Tab.~\ref{tab:models} confirms it.  Adjustments in the composition of the Universe around recombination time cannot therefore drive $H_{\rm rec}$ changes larger than that. 

 In summary, the expansion rate $H(z)$ from matter-radiation equality  through recombination and especially at recombination is very well constrained.  This tight constraint on  $H(z)$ leads in turn to constraints, at the same level, on the physical densities.
  Moreover, the amount of radiation, neutrinos and (physical) cold dark matter densities are constrained also by the perturbations. As a result, our analysis  shows the residual (small)  freedom in the early Universe composition.

The results for the ``LCDM'' case are  reported in more detail in Appendix A. These results  may be of interest beyond the scope of this paper. In fact, because  they are robust to detailed   assumptions about dark energy  properties they can be used as a dark-energy ``hardened" CMB prior for dark energy analyses.

As discussed above, one  caveat in interpreting these results is that we have excluded multipoles $\ell<30$ to remove the late-time ISW effect. In principle, for non-standard models, there could still be some late-time contribution at $\ell>30$ which, in  this approach,  could then mistakenly be interpreted as early cosmology signal.  This affects the constraints on dark energy but not on the other parameters.

\begin{figure}
\centering
\includegraphics[clip,trim=0.5cm 6cm 0.5cm 6cm,width=.65\textwidth]{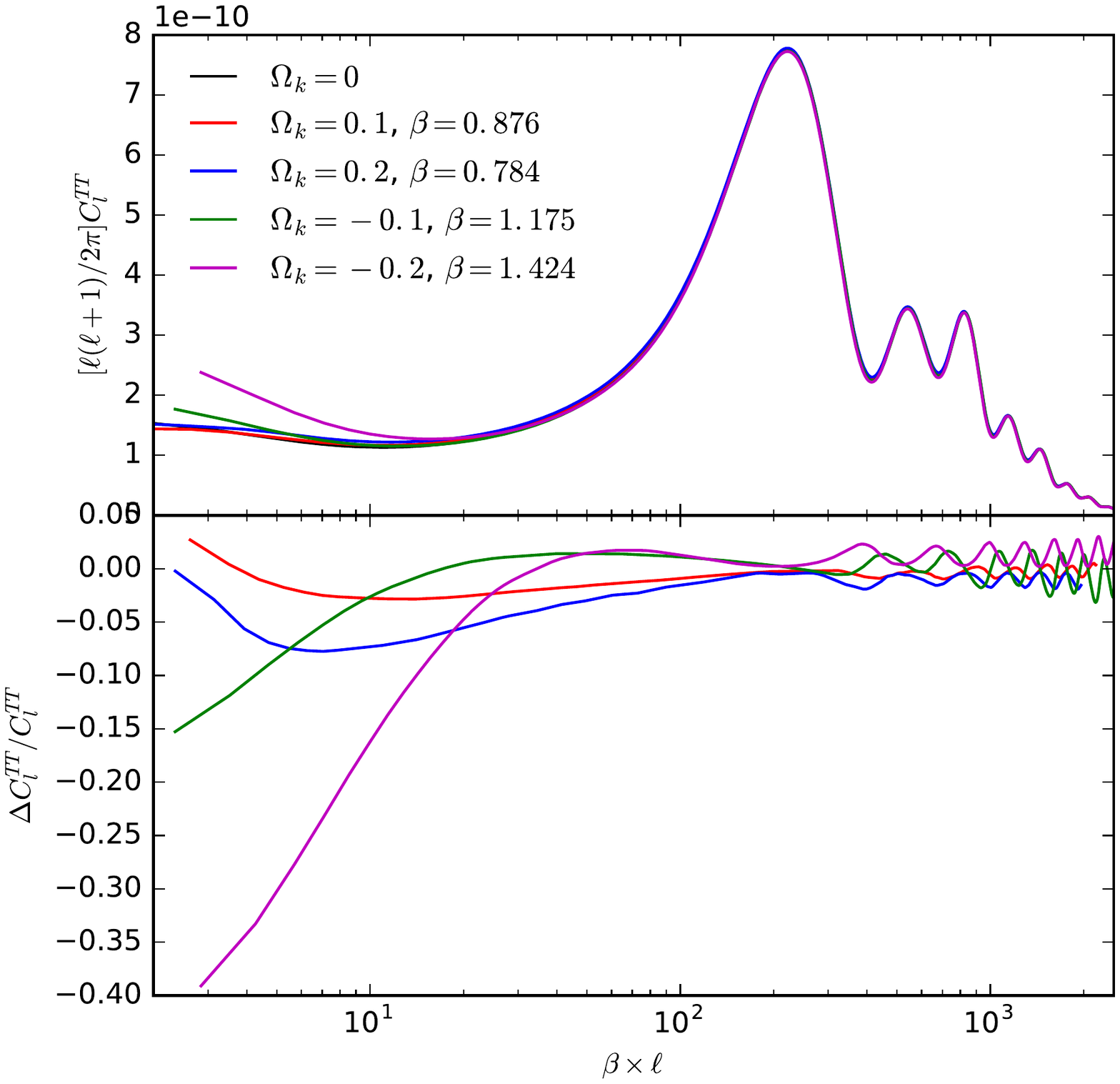}
\includegraphics[width=.65\textwidth]{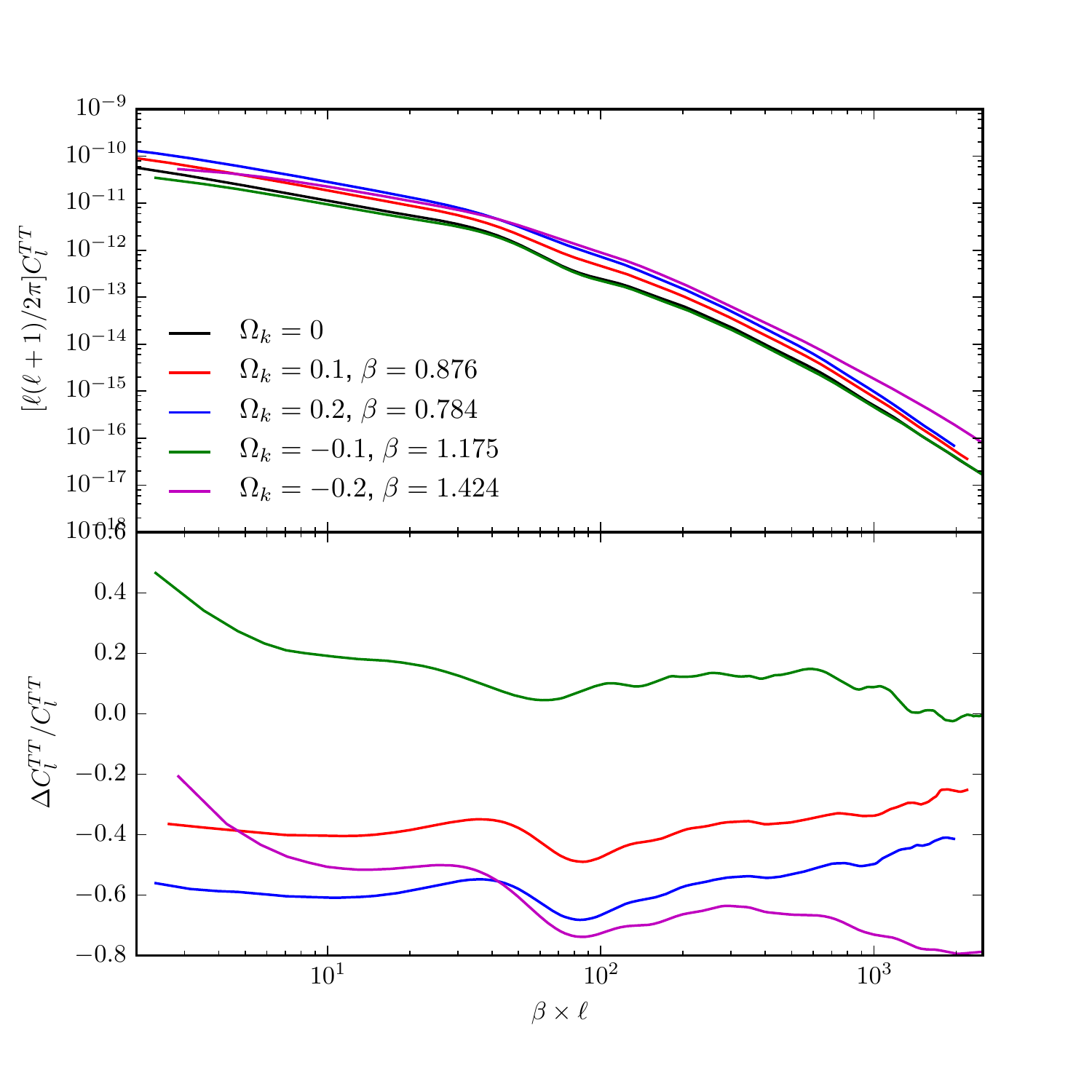}
\caption{Temperature power spectra for different non-flat  models illustrating the remaining late-time physics for current large absolute  values of the curvature (and hence values  of $\Omega_{\Lambda}$ away from the $\Lambda$CDM fiducial value). In particular, for large values of $\Omega_{\rm k}$ some  ISW signal  remains up to $\ell \gg 30$ thus modifying the shape of the first peak and rendering the $\ell$ re-scaling imperfect. The lower panels show the power spectrum of late-time ISW only. The remaining wiggles at high $\ell$ are an artefact of the imperfect $\ell$ rescaling.}
\label{fig:late}
\end{figure}

In Fig.~\ref{fig:late} we show how late-time effects that become important  when $\Omega_{\rm DE}$ at $z = 0$ is $> 0.8$ cannot be  fully removed by the approach of \cite{Vonlanthen:2010cd,Audren:2012wb}. For example, the  late ISW  effect extends up to $l \sim 200$, modifying the shape of the first peak and making the $\ell$ rescaling imperfect, when $30<\ell <200$  are included. Excluding all $\ell<200$ would yield greatly degraded constraints.

In all the plots we show also the posterior distribution of the sound-horizon at radiation drag. This is an early-time quantity which with this approach is now measured in a way that is independent of assumptions about late-time physics. The sound horizon at radiation drag, $r_{\rm s}(z_{\rm d})$, remains well constrained for all the models under consideration, with a remarkable exception. The only one that allows for a large variation of $r_{\rm s}(z_{\rm d})$ is the ``nu'' model. A possible explanation is given by the fact that this is the only model that modifies directly the properties of species (and  thus the expansion rate) that are ``naturally" important at early times (neutrinos). In the other examples, we added species that have a density that scales more rapidly than radiation in the past, and thus become progressively unimportant at  early times (e.g., $z>z_{\rm d}$).

The precision in this measurement is better than 0.5\%, which is just a factor two larger than for the standard analysis \cite{Ade:2015xua} being the central value fully consistent. It is interesting to note  that the model-independent measurement of the ``local standard ruler" \cite{Heavens16}  is in good agreement with the value of $r_{\rm d}$ found from the early-time only analysis and  has error-bars comparable to those on $r_{\rm d}$ found in the ``nu" model. As shown in Fig.~\ref{nu_posterior}, including this additional constraint slightly alters the bounds on this model, in particular it slightly reduces the error bars on $N_{\rm eff}$ bringing the 95\% credible interval to $2.4<N_{\rm eff}<3.2$. Forecasted  improvements on Baryon Acoustic Oscillation data and on measurements of $H_0$ are expected to  further reduce the  error on the ruler (see e.g., \cite{Heavens16} for a discussion), promising therefore more stringent tests on the physics in the early Universe.

\begin{table}
\centering
\begin{tabular}{|l|c|c|c|c|c|} 
 \hline
 & LCDM & DE fld & k & nu & matteriation\\
 \hline

 $10\Omega_b^{\rm rec}$ & $1.187^{+0.034}_{-0.034}$ & $1.186^{+0.034}_{-0.034}$ & $1.188^{+0.036}_{-0.034}$ & $1.205^{+0.047}_{-0.047}$ & $1.185^{+0.035}_{-0.033}$ \\

 $10\Omega_{\rm cdm}^{rec}$ & $6.378^{+0.071}_{-0.070}$ & $6.379^{+0.071}_{-0.070}$ & $6.376^{+0.071}_{-0.075}$ & $6.378^{+0.071}_{-0.072}$ & $6.351^{+0.082}_{-0.085}$ \\

 $H_{\rm rec}$ [Mpc$^{-1}$] & $5.197^{+0.045}_{-0.045}$ & $5.198^{+0.046}_{-0.045}$ & $5.196^{+0.046}_{-0.047}$ & $5.12^{+0.14}_{-0.14}$ & $5.193^{+0.046}_{-0.048}$ \\

 $n_{\rm s}$ & $0.964^{+0.010}_{-0.010}$ & $0.964^{+0.010}_{-0.010}$ & $0.964^{+0.011}_{-0.011}$ & $0.956^{+0.023}_{-0.023}$ & $0.966^{+0.011}_{-0.011}$ \\

 $e^{-2\tau} A_{\rm s}$ & $1.879^{+0.024}_{-0.025}$ & $1.879^{+0.026}_{-0.026}$ & $1.876^{+0.025}_{-0.026}$ & $1.871^{+0.072}_{-0.070}$ & $1.875^{+0.026}_{-0.026}$ \\

 \hline
 
 $10^4\Omega_{\rm k}^{\rm rec}$ & - & - & $0.27^{+1.17}_{-1.24}$ & - & - \\

 $10^3\Omega_{\rm fld}^{\rm rec}$ & - & $0.039^{+0.197}_{-0.039}$ & - & - & $2.50^{+3.40}_{-2.50}$ \\

 $w^0_{\rm fld}$ & - & $-1.15^{+0.44}_{-0.73}$ & - & - & $0.05^{+0.10}_{-0.05}$ \\

 $N_{\rm eff}$ & - & - & - & $2.77^{+0.47}_{-0.46}$ & - \\

 $c_{\rm eff}^2$ & - & - & - & $0.33^{+0.01}_{-0.01}$ & - \\

 $c_{\rm vis}^2$ & - & - & - & $0.34^{+0.10}_{-0.10}$ & - \\

\hline
 
 $z_{\rm rec }$ & $1089.0^{+0.5}_{-0.5}$ & $1089.0^{+0.5}_{-0.5}$ & $1089.0^{+0.5}_{-0.6}$ & $1088.7^{+0.6}_{-0.6}$ & $1089.0^{+0.5}_{-0.5}$ \\
 
 $r^s_{d }$ & $147.35^{+0.66}_{-0.66}$ & $147.34^{+0.66}_{-0.66}$ & $147.37^{+0.68}_{-0.69}$ & $150.0^{+4.8}_{-4.7}$ & $147.47^{+0.69}_{-0.71}$ \\

 \hline

 $\chi^2_{min}$ & $2456$ & $2454$ & $2456$ & $2454$ & $2455$ \\

\hline
\end{tabular}
\caption{Constraints on the cosmological parameters for the models considered using the TT TEEE and lensing likelihoods. The entries $w^0_{\rm fld}$ and $\Omega_{\rm fld}$ correspond to  $w^0_{\rm DE}$, $\Omega_{\rm DE}$ for the ``DE fld" model and  $w^0_{\rm MR}$,$\Omega_{\rm MR}$ for the matteriation model. Quoted limits are 95\% credible intervals.}\label{tab:models}
\end{table}

\section{Conclusions}
\label{sec:concl}

We have explored in a model-independent way, how well the early universe conditions (i.e. universe components and their properties) are constrained by current  CMB observations. To do so, we have used a method that attempts to  decouple the late-universe from the early-universe.
We have compared  the results with  model-independent determinations of the  standard ruler corresponding to the sound horizon at radiation drag from the literature, finding good agreement.

In our method, we allow for the possibility that energy-density components different from those in the $\Lambda$CDM model are present in the early Universe. In particular, we allowed for extra dark energy (``DE fld", with equation of state $w<0$), extra fluid matter (``MR", $0 < w < 1/3$), extra radiation or other relativistic species ($w=1/3$, parameterized by $N_{\rm eff}$) and extra curvature (``k''),  covering many possibilities for the expansion history of the universe at early times.   Our methodology,  as presented and implemented here works very well for models that at late time are not too drastically different from the standard $\Lambda$CDM model. In particular, there are residual late time effects when $\Omega_{\rm DE} > 0.8$ at $z \sim 0$. For dark energy constraints, the methodology as currently implemented, should be  interpreted as ``robust" to detailed assumptions about dark energy rather than fully model-independent.

 State-of-the-art CMB observations including temperature and polarization information, effectively constrain the expansion rate at recombination with $\sim 1\%$ precision. The expansion rate governs the ionization fraction and the width of the visibility function, which in turn affects the Silk damping and the amplitude of the  polarization signal \cite{ZZ}. Adjustments in the composition of the Universe  around recombination time are constrained by this effect. Moreover, components that cluster are constrained also through their effects on the perturbations.

We found that current observations constrain surprisingly tightly  these extra components: $\Omega_{\rm MR} < 0.006$ and $2.3 < N_{\rm eff} < 3.2$ when imposing spatial flatness. These energy densities are all reported as 95\% credible intervals, at $z_{\rm rec}$. For the latter case of extra radiation, when we use the local measurement of the standard ruler to further limit the amount of dark radiation, we obtain that $ 2.5 < N_{\rm eff} < 3.3 $ using SNe, BAOs and a Hubble prior (SBH), and $2.4 < N_{\rm eff} < 3.2$ replacing the Hubble prior with cosmic chronometers (CSB).

Our conclusion is that current  CMB (temperature and polarization) observations, alone or  in conjunction with low-redshift, model-independent  measurements of the standard ruler,  tell us that the early-universe (up to recombination) is very well known: the data require the presence of baryons,  radiation,  a fluid that behaves like neutrinos and dark matter; dark energy and curvature are negligible. There is  no evidence --and not much room-- for exotic  matter-energy components besides those in the $\Lambda$CDM; the  standard $\Lambda$CDM model describes extremely well the early cosmology.

 We envision that  model-independent  approaches like the one presented here will be of value as cosmology moves beyond parameter fitting and towards model testing.

\section*{Acknowledgements}
Funding for this work was partially provided by the Spanish MINECO under projects  AYA2014-58747-P and MDM-2014- 0369 of ICCUB  (Unidad de Excelencia Maria de Maeztu), Royal Society grant IE140357 and by CNPq (Brazil).  We acknowledge the support of Imperial College through the CosmoCLASSIC collaboration. 

\section*{Appendix A}

Large-scale structure, supernovae or other low redshift observations analyses on dark energy properties are often slowed down significantly by the fact that,  to include CMB  information, a Boltzmann code must be suitably  modified and ran  for each specific model. Not including CMB information simplifies the analysis greatly, but also reduces dramatically the constraints.

One approach adopted so far has been to use the so-called CMB distance priors (\cite{Wang:2007mza,Li:2008cj} and \cite{Huang:2015vpa} and references therein).

Here we propose a slightly more sophisticated approach of using the results provided by the (late-time model independent) ``LCDM" model analysis and reported in Tab. \ref{tab:models}.  These results are by construction independent of late-time assumptions about  perturbations and are ``robust" to detailed assumptions about  the expansion history  and in particular about the detailed dark energy properties, yet they fully capture the CMB information assuming standard early-time physics. The values for the density parameters are reported at the redshift of recombination. They can then be easily extrapolated to lower redshifts for any given model for the Universe expansion history.  
The recombination redshift is well constrained and does not show significant degeneracy with any of the parameters, so for this purpose it can be assumed  to be fixed at $z_{\rm rec}=1089$.

%%\bibliographystyle{utcaps}  %this sytle wasn't working (or I probably don't know the correct command)
%\bibliographystyle{unsrt} %in order of appearance 
%%\bibliographystyle{plainnat} % alphabetic order
%\bibliography{biblio}

\begin{thebibliography}{10}

\bibitem{Vonlanthen:2010cd}
Marc Vonlanthen, Syksy Rasanen, and Ruth Durrer,
\newblock {Model-independent cosmological constraints from the CMB},
\newblock {\em JCAP}, 1008:023, 2010.

\bibitem{Audren:2012wb}
Benjamin Audren, Julien Lesgourgues, Karim Benabed, and Simon Prunet,
\newblock {Conservative Constraints on Early Cosmology: an illustration of the
  Monte Python cosmological parameter inference code},
\newblock {\em JCAP}, 1302:001, 2013.

\bibitem{Audren:2013nwa}
Benjamin Audren.
\newblock {Separate Constraints on Early and Late Cosmology},
\newblock {\em Mon. Not. Roy. Astron. Soc.}, 444(1):827--832, 2014,

\bibitem{Heavens:2014rja}
Alan Heavens, Raul Jimenez, and Licia Verde,
\newblock {Standard rulers, candles, and clocks from the low-redshift
  Universe},
\newblock {\em Phys. Rev. Lett.}, 113(24):241302, 2014.

\bibitem{Sutherland:2012ys}
Will Sutherland,
\newblock {On measuring the absolute scale of baryon acoustic oscillations},
\newblock {\em Mon. Not. Roy. Astron. Soc.}, 426:1280, 2012.


%\cite{Adam:2015rua}
\bibitem{Adam:2015rua}
  R.~Adam {\it et al.} [Planck Collaboration],
  Planck 2015 results. I. Overview of products and scientific results,
  Astron.\ Astrophys.\  {\bf 594} (2016) A1
  doi:10.1051/0004-6361/201527101
  [arXiv:1502.01582 [astro-ph.CO]].
  %%CITATION = doi:10.1051/0004-6361/201527101;%%
  %309 citations counted in INSPIRE as of 21 Oct 2016

%\cite{Aghanim:2015xee}
\bibitem{Aghanim:2015xee}
  N.~Aghanim {\it et al.} [Planck Collaboration],
  Planck 2015 results. XI. CMB power spectra, likelihoods, and robustness of parameters,
  Astron.\ Astrophys.\  {\bf 594} (2016) A11
  doi:10.1051/0004-6361/201526926
  [arXiv:1507.02704 [astro-ph.CO]].
  %%CITATION = doi:10.1051/0004-6361/201526926;%%
  %185 citations counted in INSPIRE as of 21 Oct 2016

%\cite{Ade:2015xua}
\bibitem{Ade:2015xua}
  P.~A.~R.~Ade {\it et al.} [Planck Collaboration],
  Planck 2015 results. XIII. Cosmological parameters,
  Astron.\ Astrophys.\  {\bf 594} (2016) A13
  doi:10.1051/0004-6361/201525830
  [arXiv:1502.01589 [astro-ph.CO]].
  %%CITATION = doi:10.1051/0004-6361/201525830;%%
  %2302 citations counted in INSPIRE as of 21 Oct 2016





%\cite{Lesgourgues:2011re}
\bibitem{2011arXiv1104.2932L}
  J.~Lesgourgues,
  ``The Cosmic Linear Anisotropy Solving System (CLASS) I: Overview,''
  arXiv:1104.2932 [astro-ph.IM].
  %%CITATION = ARXIV:1104.2932;%%
  %141 citations counted in INSPIRE as of 21 Oct 2016

\bibitem{Blas:2011rf}
Diego Blas, Julien Lesgourgues, and Thomas Tram,
\newblock {The Cosmic Linear Anisotropy Solving System (CLASS) II:
  Approximation schemes},
\newblock {\em JCAP}, 1107:034, 2011.

%\cite{Verde:2016ccp}
\bibitem{Heavens16}
  L.~Verde, J.~L.~Bernal, A.~F.~Heavens and R.~Jimenez,
 \newblock{The length of the low-redshift standard ruler},
  \newblock{arXiv:1607.05297 [astro-ph.CO]}.
  %%CITATION = ARXIV:1607.05297;%%

\bibitem{Cuesta:2015mqa}
Antonio~J. Cuesta et~al,
\newblock {The clustering of galaxies in the SDSS-III Baryon Oscillation
  Spectroscopic Survey: Baryon Acoustic Oscillations in the correlation
  function of LOWZ and CMASS galaxies in Data Release 12},
\newblock {\em Mon. Not. Roy. Astron. Soc.}, 457:1770, 2016.

\bibitem{Betoule:2014frx}
M.~Betoule et~al,
\newblock {Improved cosmological constraints from a joint analysis of the
  SDSS-II and SNLS supernova samples},
\newblock {\em Astron. Astrophys.}, 568:A22, 2014.

\bibitem{Kazin:2014qga}
Eyal~A. Kazin et~al,
\newblock {The WiggleZ Dark Energy Survey: improved distance measurements to z
  = 1 with reconstruction of the baryonic acoustic feature},
\newblock {\em Mon. Not. Roy. Astron. Soc.}, 441(4):3524--3542, 2014.

\bibitem{Beutler:2011hx}
Florian Beutler, Chris Blake, Matthew Colless, D.~Heath Jones, Lister
  Staveley-Smith, Lachlan Campbell, Quentin Parker, Will Saunders, and Fred
  Watson.
\newblock {The 6dF Galaxy Survey: Baryon Acoustic Oscillations and the Local
  Hubble Constant}.
\newblock {\em Mon. Not. Roy. Astron. Soc.}, 416:3017--3032, 2011.


%\cite{Riess:2016jrr}
\bibitem{Riess:2016jrr}
  A.~G.~Riess {\it et al.},
  A 2.4\% Determination of the Local Value of the Hubble Constant,
  Astrophys.\ J.\  {\bf 826} (2016) no.1,  56
  doi:10.3847/0004-637X/826/1/56
  [arXiv:1604.01424 [astro-ph.CO]].
  %%CITATION = doi:10.3847/0004-637X/826/1/56;%%
  %93 citations counted in INSPIRE as of 21 Oct 2016

\bibitem{Moresco:2016mzx}
Michele Moresco, Lucia Pozzetti, Andrea Cimatti, Raul Jimenez, Claudia
  Maraston, Licia Verde, Daniel Thomas, Annalisa Citro, Rita Tojeiro, and David
  Wilkinson,
\newblock {A 6\% measurement of the Hubble parameter at $z\sim0.45$: direct
  evidence of the epoch of cosmic re-acceleration},
\newblock 2016.

\bibitem{Wetterich:2004pv}
Christof Wetterich,
\newblock {Phenomenological parameterization of quintessence},
\newblock {\em Phys. Lett.}, B594:17--22, 2004.

\bibitem{Doran:2006kp}
Michael Doran and Georg Robbers,
\newblock {Early dark energy cosmologies},
\newblock {\em JCAP}, 0606:026, 2006.

%\cite{Ade:2013sjv}
\bibitem{Ade:2013sjv}
  P.~A.~R.~Ade {\it et al.} [Planck Collaboration],
  Planck 2013 results. I. Overview of products and scientific results,
  Astron.\ Astrophys.\  {\bf 571} (2014) A1
  doi:10.1051/0004-6361/201321529
  [arXiv:1303.5062 [astro-ph.CO]].
  %%CITATION = doi:10.1051/0004-6361/201321529;%%
  %993 citations counted in INSPIRE as of 21 Oct 2016

%\cite{Ade:2015rim}
\bibitem{Ade:2015rim}
  P.~A.~R.~Ade {\it et al.} [Planck Collaboration],
  Planck 2015 results. XIV. Dark energy and modified gravity,
  Astron.\ Astrophys.\  {\bf 594} (2016) A14
  doi:10.1051/0004-6361/201525814
  [arXiv:1502.01590 [astro-ph.CO]].
  %%CITATION = doi:10.1051/0004-6361/201525814;%%
  %197 citations counted in INSPIRE as of 21 Oct 2016


\bibitem{Lewis:1999bs}
Antony Lewis, Anthony Challinor, and Anthony Lasenby,
\newblock {Efficient computation of CMB anisotropies in closed FRW models},
\newblock {\em Astrophys. J.}, 538:473--476, 2000.

 
\bibitem{Ballesteros:2010ks}
Guillermo Ballesteros and Julien Lesgourgues.
\newblock {Dark energy with non-adiabatic sound speed: initial conditions and
  detectability},
\newblock {\em JCAP}, 1010:014, 2010.

\bibitem{KMJ}
  A.~Kosowsky, M.~Milosavljevic and R.~Jimenez,
  Efficient cosmological parameter estimation from microwave background anisotropies,
  Phys.\ Rev.\ D {\bf 66} (2002) 063007
  doi:10.1103/PhysRevD.66.063007
  [astro-ph/0206014].



%\cite{Zumalacarregui:2016pph}
\bibitem{Zumalacarregui:2016pph}
  M.~Zumalacrregui, E.~Bellini, I.~Sawicki and J.~Lesgourgues,
  hiclass: Horndeski in the Cosmic Linear Anisotropy Solving System,
  arXiv:1605.06102 [astro-ph.CO].
  %%CITATION = ARXIV:1605.06102;%%
  %11 citations counted in INSPIRE as of 21 Oct 2016


\bibitem{Bellini:2014fua}
Emilio Bellini and Ignacy Sawicki,
\newblock {Maximal freedom at minimum cost: linear large-scale structure in
  general modifications of gravity},
\newblock {\em JCAP}, 1407:050, 2014.

\bibitem{BelliniSawicki}
E.~{Bellini} and I.~{Sawicki},
\newblock {Maximal freedom at minimum cost: linear large-scale structure in
  general modifications of gravity},
\newblock {\em JCAP}, 7:050, July 2014.

\bibitem{Hu:1998tk}
Wayne Hu, Daniel~J. Eisenstein, Max Tegmark, and Martin~J. White,
\newblock {Observationally determining the properties of dark matter},
\newblock {\em Phys. Rev.}, D59:023512, 1999.

\bibitem{Karwal:2016vyq}
  T.~Karwal and M.~Kamionkowski,
  Early dark energy, the Hubble-parameter tension, and the string axiverse,
  arXiv:1608.01309 [astro-ph.CO].
  %%CITATION = ARXIV:1608.01309;%%
  
\bibitem{ZZ}
O.~{Zahn} and M.~{Zaldarriaga},
\newblock {Probing the Friedmann equation during recombination with future
  cosmic microwave background experiments},
\newblock {\em \it Phys. Rev. Lett}, 67(6):063002, 2003.

\bibitem{Wang:2007mza}
Yun Wang and Pia Mukherjee,
\newblock {Observational Constraints on Dark Energy and Cosmic Curvature},
\newblock {\em Phys. Rev.}, D76:103533, 2007.

\bibitem{Li:2008cj}
Hong Li, Jun-Qing Xia, Gong-Bo Zhao, Zu-Hui Fan, and Xinmin Zhang,
\newblock {On using the WMAP distance priors in constraining the time evolving
  equation of state of dark energy},
\newblock {\em Astrophys. J.}, 683:L1--L4, 2008.

\bibitem{Huang:2015vpa}
Qing-Guo Huang, Ke~Wang, and Sai Wang,
\newblock {Distance priors from Planck 2015 data},
\newblock {\em JCAP}, 1512(12):022, 2015.

\end{thebibliography}

\end{document}